\documentclass[print]{aa} 
\usepackage{savesym}
\usepackage{graphicx}
\usepackage{footnote}
\usepackage{hyperref}
\usepackage{lineno}
\usepackage{mathtools}
\usepackage{xspace}
\usepackage{units}
\usepackage{relsize}
\usepackage{tikz}
\usepackage{amssymb,amsfonts,amsmath,amstext,amsgen,amsopn,amsxtra,indentfirst,times}
\usetikzlibrary{positioning}
\usetikzlibrary{shapes,arrows}
\usepackage{multirow}
\usepackage{nicefrac}
\usepackage{array}
\usepackage{wasysym}

\DeclareFontFamily{U}{mathb}{\hyphenchar\font45}
\DeclareFontShape{U}{mathb}{m}{n}{
      <5> <6> <7> <8> <9> <10> gen * mathb
      <10.95> mathb10 <12> <14.4> <17.28> <20.74> <24.88> mathb12
      }{}
\DeclareSymbolFont{mathb}{U}{mathb}{m}{n}
\DeclareMathSymbol{\Earth}{3}{mathb}{"43}

\def\widesplit#1{%
\cleardoublepage
\def\row##1##2{##1}%
#1%
\\

\def\row##1##2{##2}%
#1%
\clearpage
}
\newcommand{\coco}{\color{black}}
\newcommand{\mice}{$m_{\rm water}$\xspace}

\newcommand{\RE}{R$_{\Earth}$\xspace}
\newcommand{\ME}{M$_{\Earth}$\xspace}
\newcommand{\RS}{R$_{\rm sun}$\xspace}
\newcommand{\Zenv}{$Z_{\rm env}$\xspace}
\newcommand{\rc}{$r_{\rm c}$\xspace}
\newcommand{\rsolid}{$r_{\rm mantle}$\xspace}
\newcommand{\menv}{$m_{\rm env}$\xspace}

\newcommand{\m}{{\bf m}\xspace}
\newcommand{\dat}{{\bf d}\xspace}

\newcommand{\fesi}{{\rm Fe}/{\rm Si}_{\rm bulk}\xspace}
\newcommand{\mgsi}{{\rm Mg}/{\rm Si}_{\rm bulk}\xspace}
\newcommand{\fesistar}{{\rm Fe}/{\rm Si}_{\rm star}\xspace}
\newcommand{\mgsistar}{{\rm Mg}/{\rm Si}_{\rm star}\xspace}
\newcommand{\fesima}{{\rm Fe}/{\rm Si}_{\rm mantle}\xspace}
\newcommand{\mgsima}{{\rm Mg}/{\rm Si}_{\rm mantle}\xspace}

\begin{document}

   \title{ Bayesian analysis of interiors of \mbox{HD~219134b}, Kepler-10b, Kepler-93b, CoRoT-7b, 55~Cnc~e, and \mbox{HD~97658b} using stellar abundance proxies}
  % Using stellar abundance proxies to constrain the structure of \mbox{HD~219134b}, Kepler-10b, Kepler-93b, CoRoT-7b, 55~Cnc~e, and \mbox{HD~97658b}}
   \author{Caroline Dorn\inst{1}
              \and Natalie R. Hinkel\inst{2}
              \and Julia Venturini\inst{1}}
 \authorrunning{Dorn et al.}
  \titlerunning{Probabilistic analysis of planet interiors using stellar abundance proxies}
  
            \institute{Physics Institute, University of Bern, Sidlerstrasse 5, CH-3012, Bern, Switzerland\\
              \email{caroline.dorn@space.unibe.ch}
         \and
             School of Earth \& Space Exploration, Arizona State University,
  Tempe, AZ 85287, USA \\
             }

\abstract%{}{}{}{}{} 
% 5 {} token are mandatory
  {}
   {Using a generalized Bayesian inference method, we aim to explore the possible interior structures of six selected exoplanets for which planetary mass and radius measurements are available in addition to stellar host abundances: HD~219134b, Kepler-10b, Kepler-93b, CoRoT-7b, 55~Cnc~e, and HD~97658b. We aim to investigate the importance of stellar abundance proxies for the planetary bulk composition (namely Fe/Si and Mg/Si) on prediction of planetary interiors.}
   {We performed a full probabilistic Bayesian inference analysis to formally account for observational and model uncertainties while obtaining confidence regions of structural and compositional parameters of core, mantle, ice layer, ocean, and atmosphere. We determined how sensitive our parameter predictions depend on (1) different estimates of bulk abundance constraints and (2) different correlations of bulk abundances between planet and host star.}
   { The possible interior structures and  correlations between structural parameters differ depending on data and data uncertainty. The strongest correlation is generally found between size of rocky interior and water mass fraction. Given the data, possible water mass fractions are  high, even for most potentially rocky planets (\mbox{HD~219134b}, Kepler-93b, CoRoT-7b, and 55~Cnc~e with estimates up to 35~\%, depending on the planet). Also, the interior of Kepler-10b is best constrained with possible interiors similar to Earth. Among all tested planets, only the data of Kepler-10b and Kepler-93b allow to put a higher probability on the planetary bulk Fe/Si to be stellar compared to extremely sub-stellar.}
{Although the possible ranges of interior structures are large, structural parameters and their correlations are constrained by the sparse data. The probability for the tested exoplanets to be Earth-like is generally very low. Furthermore, we conclude that different estimates of planet bulk abundance constraints mainly affect mantle composition and core size.}


\maketitle

%%%%%%%%%%%%%%%%%%%%%%%%%%%%%%%%%%%%%%%%%%%%%%%%%%%%%%%%%%%%%%%%%%%%%%%%%%%%%%
\section{Introduction}
%\begin{linenumbers}
\label{Intro}
%\begin{linenumbers}
%%
\sloppy
The characterization of exoplanet interiors is key to understand planet diversity. Here we focus on characterizing six exoplanets for which mass, radius, and also refractory element abundances of the stellar hosts are known. These exoplanets are  \mbox{HD~219134b}, Kepler-10b, Kepler-93b, CoRoT-7b, 55~Cnc~e, and \mbox{HD~97658b}. In order to make meaningful statements about their interior structures, it is mandatory not only to find few interior realizations that explain the data, but also to quantify the spread in possible interior structures that are in agreement with data, and data and model uncertainty \citep{rogers2010,dorn,dornA}. This spread is generally large, due to the inherent degeneracy, i.e. different interior realizations can have identical mass and radius. We therefore apply the methodology presented by \citet{dornA} that employs a full probabilistic Bayesian inference analysis using a Markov chain Monte Carlo (McMC) method. This method is validated against Neptune and previously by \citet{dorn} against the terrestrial fitsyno planets. It allows us to determine confidence regions for structural and compositional parameters of core, mantle, ice layer, ocean, and atmosphere.

Planet bulk abundance constraints, in addition to mass and radius, are crucial to reduce the generally high degeneracy in interior structures \citep{dorn,grasset09}. Stellar abundances in relative refractory elements (Fe/Si and Mg/Si) were suggested to serve as a proxy for the planetary abundance \citep[e.g.,][]{bond,elser,johnson,thiabaud}. Spectroscopic observations of relative photospheric abundances of host stars may allow us to determine their importance for the interior characterization of planetary companions as outlined by \citet{santos}. We proceed along these lines and study the sensitivity of interior parameter predictions to different planet bulk abundances by testing (1) different stellar abundance estimates and (2) different correlations between stellar and planetary abundance ratios.

Over the last few decades, a variety of telescopes and spectrographs have been utilized to measure stellar abundances, which affects the resolution and signal-to-noise (S/N) ratios of the data. In addition, the techniques used to determine the element abundances differ. Then, each research group chooses their own line list and number of ionization stages included, solar atmosphere models, and adopted solar abundances. All of these differences result in element abundance measurements that are fundamentally on different baselines and which do not overlap to within error (Hinkel et al. 2014). Here, we investigate whether these discrepancies significantly affects the determination of interior structure or not.

The six selected exoplanets for which host star abundances are available have bulk densities in the  range from 6.8 g/cm$^{3}$ (Kepler-93b) to 3.4 g/cm$^{3}$ (HD~97658b) and span across the proposed transition range between mostly rocky and mostly non-rocky exoplanets \citep[e.g.,][]{lopez,rogers15}. Terrestrial-type planets are generally thought to be differentiated with an iron-rich core, a
silicate mantle, and a crust, whereas volatile-rich planets are thought to also contain significant amount of ices and/or atmospheres \citep[e.g., see][]{howe}. Direct atmosphere characterization for the selected planets are not available, except for the only recently characterized atmosphere of 55~Cnc~e \citep{Tsiaras, demory}. 
For super Earths in general, the diversity of atmospheric structures and compositions are mostly theoretical \citep[e.g., see][for a review]{burrows}.
Studies of mass-radius relations generally consider H$_{2}$-He atmospheres \citep[e.g.,][]{rogers2011, fortney} and H$_2$O as liquid and high pressure ice \citep[e.g.,][]{valencia07a, seager2007}.  However, compositional diversity of atmospheres and ices in planets may exceed the one found in our solar system \citep[e.g.,][]{newsom}. But the sparse observational data on exoplanets will only allow us to constrain few structural and compositional parameters. In that light, we assume a general planetary structure of a pure iron core, a silicate mantle, a water layer, and an atmosphere made of H, He, C, and O. Given the data, we constrain core size, mantle thickness and composition, mass of water, and key characteristics of the atmosphere (mass fraction, luminosity, and metallicity). Here, we use the term atmosphere	synonymously with gas layer, for which there is commonly a distinction between a convective (i.e., envelope) and a radiative part (i.e., atmosphere).

In order to compute the structure for a planet of a given interior, we employ model I from \citet[][]{dornA}. Namely, we employ state-of-the-art structural models that compute irradiated atmospheres and self-consistent thermodynamics of core, mantle, high-pressure ice and ocean.

This paper is organized as follows: in Section \ref{prev} we review previous interior studies on the planets of interest. In Section \ref{Methodology} we introduce the inference strategy, model parameters, data and the
physical model that links data and model parameters. In Section \ref{Results}, we apply our method on
the six selected exoplanets and test different stellar abundance proxies for constraints on planet bulk abundance. We end with discussion and conclusions. 

 \begin{figure}[]
\centering
 \includegraphics[width = .5\textwidth]{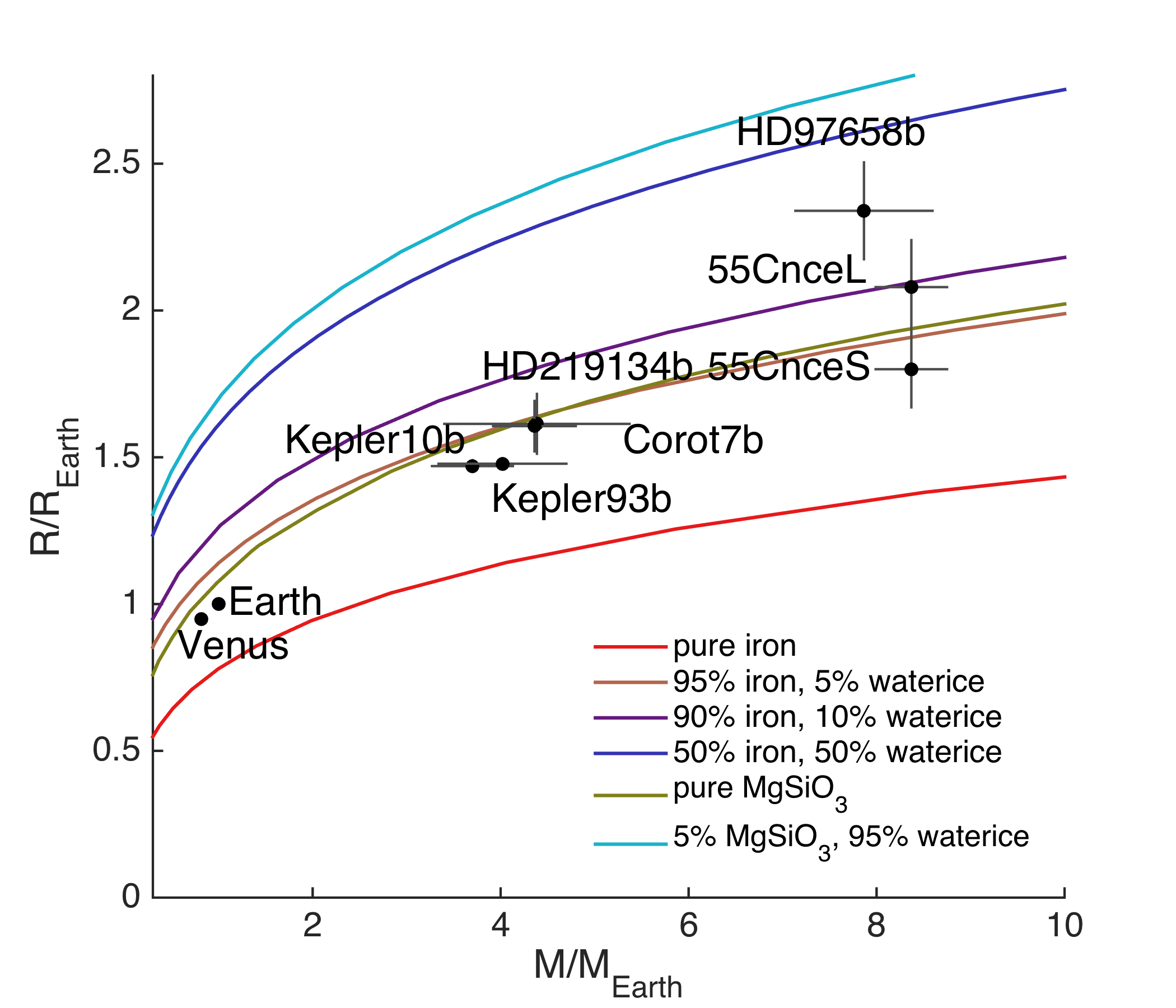}
 \caption{Masses and radii of the six exoplanets of interest, including two scenarios for 55Cnc e (see main text). The terrestrial solar system planets are also plotted against idealized mass-radius curves of pure iron, pure magnesium silicate, and a 2-layered composition of 5~\% magnesium silicate and 95~\% water. \label{MR}}
\end{figure}

\section{Previous studies of the selected exoplanets}
\label{prev}
 Previous estimates on the interiors of HD~219134b, Kepler-10b, Kepler-93b, CoRoT-7b, 55~Cnc~e, and HD~97658b generally do not comprehensively reflect the inherent degeneracy that different interiors can explain the observations. Often a few interior scenarios are discussed but it remains unclear how well specific models compare with the generally large number of possible scenarios.
Commonly used mass-radius curves of idealized compositions plotted against the planetary data (Figure \ref{MR}) allow a general evaluation whether a planet can either be composed of pure rocks or is volatile-rich. 

For Kepler-10b, \citet{Batalha} estimate a predominantly rocky composition of iron and silicates, with a mass fraction of water that can be as high as 25~\%. With the additional constraint of maximum mantle stripping during planet formation \citep{Marcus2009}, they predict a dry iron-rich interior. For both Kepler-10b and CoRoT-7b, \citet{Wagner} argue for a large iron core similar to Mercury ($\sim$60~\% iron and $\sim$30~\% silicates), whereas \citet{dressing, weiss} suggest an Earth-like interior (17~\% iron and 83~\% silicates). The latter conclusion also holds for Kepler-93b \citep{dressing}. For Kepler-10b and CoRoT-7b it remains unclear if they may or may not be remnants of larger volatile-rich planets that have lost their atmospheres \citep[e.g.,][]{kuro, valencia10}. 

We find that HD~219134b is very similar in size and mass to CoRoT-7b. Compared to CoRoT-7b, HD~219134b has higher precision on mass and radius and it receives less energy from its host star. Given a bulk  density of 5.76~g/cm$^3$, \citet{motalebi} suggest a rocky composition and describe a best fit two-component model of iron-magnesium silicate mantle (77~\%) and a core mass fraction of approximately 23~\%.

The bulk density of 55~Cnc~e (R = 2.0~\RE) is too low to be made of pure rocks and needs to contain significant mass of water, gas, or other light elements \citep{winn}, although \citet{gillon} argue that any kind of atmosphere is unlikely due to the intense stellar irradiation. Previous estimates of water mass fractions that can explain the data are e.g., $\sim$~45~\% \citep{endl} or $\sim$ 20~\% \citep{gillon}. We note that both estimates are based on differently assumed bulk densities and rock/silicate partitions. Recent infrared monitoring revealed a high surface temperature difference between day and night side which implies either an optically thick atmosphere with heat recirculation confined to the planetary dayside or no atmosphere \citep{demory16}. Also, recent spectroscopic observations have suggested a hydrogen-rich atmosphere \citep{Tsiaras}. 
For 55~Cnc~e, different radii can be found in the literature \citep[][and references therein]{demory}.  Here, we simply distinguish two cases 55~Cnc~e$^{\rm L}$ (R = 2.08~\RE) and 55~Cnc~e$^{\rm S}$ (R~=~1.75~\RE), where the superscripts refer to small (S) and large (L), both of which can be seen in Figure \ref{MR}.

The least dense and irradiated exoplanet in this study, HD~97658b, can be explained by a rocky core of at least 60~\% in mass with an envelope of lighter elements and little H-He components, at most 2~\% by mass \citep{vangrootel}. But its density is also consistent with a water-dominated planet \citep{dragomir}.

%%
%%%%%%%%%%%%%%%%%%%%%%%%%%%%%%%%%%%%%%%%%%%%%%%%%%%%%%%%%%%%%%%%%%%%%%%%%%%%%%
\section{Methodology}
\label{Methodology}
We use the methodology of \citet[][model I]{dornA} that determines possible ranges of interiors which can include thin and thick atmospheres. We outline the main aspects of the method in the following.

\begin{figure}[]
\centering
 \includegraphics[width = .5\textwidth]{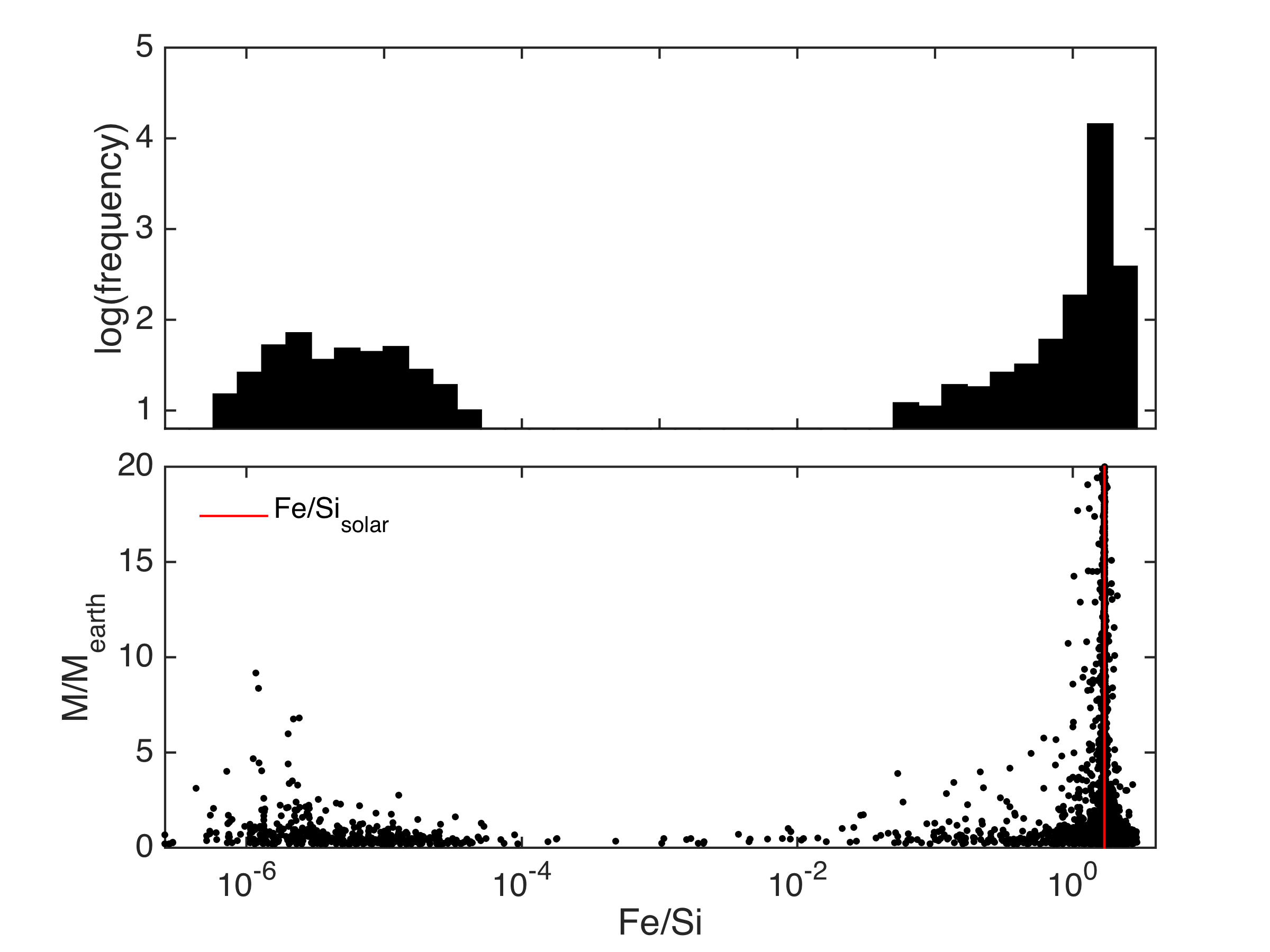}
 \caption{Top: Histogram of bulk Fe/Si of $\sim$ 15400 simulated planets that accreted from discs of solar composition. Bottom: planet mass over bulk Fe/Si of the same simulations (10 planet embryos were used in each simulation). {  Note the log-scale.} The distribution of planets is a result from \citet[][]{thiabaud}.  \label{hist}}

\end{figure}

\subsection{Bayesian inference}
\label{inversion}

We employ a Bayesian method to compute the posterior probability density function (pdf) of the model parameter conditional on the data and prior information. The prior probability of the model represents the information known about the interior structure before using data.
According to Bayes' theorem, the posterior distribution for a fixed model parametrization, conditional on data, is proportional to prior information on model parameters and the likelihood function. The likelihood function summarizes the distance (typically a norm of a vector of residuals) between the model simulation and observed data. It is practically impossible to analytically derive the posterior distribution since the addressed problem is high-dimensional and non-linear and data are sparse. We therefore employ a McMC sampling method that iteratively searches the prior parameter space and locally evaluates the likelihood function. Samples of model parameters that are generated with this approach are distributed
according to the underlying posterior distribution. As in \citet{dornA}, we use the Metropolis-Hastings algorithm to efficiently explore the posterior distribution. We refer to \citet[][]{dorn} and \citet[][]{dornA} for more details.

The large number of models that is needed for the analysis requires very efficient computations. In our work, generating the planet's internal structure takes on average 40 - 90 seconds of CPU time on a four quad-core AMD Opteron 8380 CPU node and 32 GB of RAM. In all, we sampled about $10^6$ models and retained around $5\times10^4$ models for further analysis. 

\subsection{Bulk abundance constraints}
\label{bulky}

Theoretical and empirical evidences suggest that relative abundances of refractory elements are correlated between star and planet.
For example, there is a high similarity among Fe/Si and Mg/Si for Sun, Earth, Mars, the Moon as well as meteorites \citep{lodders03,drake,mcdono,khan08,kuskov}, which are believed to represent building blocks of planets \citep{morgan}. 
Planet formation studies have demonstrated that there are direct correlations among relative refractory element abundances (e.g., Fe, Mg, and Si) between star and hosted planets \citep[e.g.,][]{bond,elser,johnson,thiabaud}. This is because refractory elements Fe, Si, and Mg condense at similar temperatures corresponding to small distances from the host star ($\gtrsim$ 0.06 AU). 

\citet{thiabaud} verified the direct correlation for a wide range of stellar compositions. For {  their tested} solar-like disk composition (e.g., Fe/Si$_{\rm Sun}$ = 1.69, Mg/Si$_{\rm Sun}$ = 0.89), the full distribution of planetary bulk Fe/Si is shown in Figure \ref{hist}. For {  the subset} of planets with semi-major axes below 0.1 AU, the distribution is {  very} weakly bimodal with 8~\% of all planets having a very low iron content (Fe/Si~$<0.001$). For this {  particular} solar-like disk composition, the authors showed that at close distances to the star (0 -- 0.2 AU), the disk's Fe/Si ($10^{-6}$--$10^{-4}$) and Mg/Si (0.08--4) mass ratios can significantly differ from solar values \citep[see also Figure 3 in][]{thiabaud}. This is because the condensation conditions differ among Mg-Si-Fe-bearing species. {  This seems to indicate that a small fraction of planets that accrete material from respective regions may have very low iron content. Here, we test the sensitivity of our interior predictions on such extremely different stellar abundance proxies for the planetary bulk composition} (see section \ref{nondirect}):
\begin{itemize}
    \item direct abundance proxy: \\$\fesi = \fesistar$\\
    $\mgsi = \mgsistar$\\
    \item non-direct abundance proxy:\\$\fesi = 10^{-5}~ \fesistar$\\
    $\mgsi = 0.5 ~\mgsistar$\\
\end{itemize}
{ We note that \citet{thiabaud} does neglect the drift of planetesimals, which may lead to an underestimation of bulk Fe/Si in their simulated planets. Future studies need to improve our understanding of the evolution of element ratios.}

\subsection{Stellar abundance data}
\label{abundancedata}

We have culled stellar abundances from multiple datasets for HD~219134, 55~Cnc~e, and HD~97658b and homogenized them via the Hypatia Catalog (Hinkel et al. 2014). In order to properly compare the datasets with each other, we have renormalized the abundances to the Lodders et al. (2009) solar abundance scale. To maintain {  coherence} between abundance measurements, our analysis did not take into consideration groups who utilized NLTE (non-local thermodynamic equilibrium) approximations. The different stellar abundance estimates of 55~Cnc~e, HD~97658, and HD~219134 are listed in Table \ref{stardata}. For Kepler-10 b, Kepler-93 b, and CoRoT-7 b, the stellar abundances have only been determined by one group, that is Santos et al. (2015).

Note, ratios with squared brackets ([X/H], e.g., Table \ref{stardata})  are logarithmic particle ratios in reference to the Sun. The majority of our discussion is with respect to abundance mass ratios without reference to the Sun (X/H, e.g., Table \ref{tabledata}).

\setlength{\tabcolsep}{6pt}
\begin{table}[ht]
\caption{Stellar abundances. The unit is dex (see Appendix \ref{app}). \label{stardata}}
\begin{center}
\begin{tabular}{llll}
\hline\noalign{\smallskip}
parameter & HD~219134 &   55 Cnc & HD~97658 \\
\hline\noalign{\smallskip}
$[\rm Fe/H]$ & 0.04--0.84& 0.29-- 0.67& -0.05~--~-0.28\\
\vspace{2mm}
$[\rm Fe/H]_{median}$ & 0.13  & 0.44 & -0.13 \\
$[\rm Mg/H]$ & 0.09--0.37& 0.34 --0.80 & -0.18~--~-0.27  \\
\vspace{2mm}
$[\rm Mg/H]_{median}$ & 0.32 & 0.41 & -0.22 \\
$[\rm Si/H]$& 0.04--0.27& 0.29 --0.57 & -0.20~--~-0.28 \\
\vspace{2mm}
$[\rm Si/H]_{median}$ &0.12 & 0.40& -0.21\\
$[\rm Na/H]$& 0.17--0.32 & 0.23-- 0.63& -0.24~--~-0.26 \\
\vspace{2mm}
$[\rm Na/H]_{median}$&0.19& 0.49&-0.25\\
$[\rm Al/H]$& 0.16--0.29& 0.43--0.55& -0.23  \\
\vspace{2mm}
$[\rm Al/H]_{median}$& 0.23& 0.51 & -0.23\\
$[\rm Ca/H]$& 0.18--0.25& 0.09 --0.46& -0.25\\
$[\rm Ca/H]_{median}$ & 0.21& 0.24& -0.25\\
\hline
\end{tabular}
\end{center}
\end{table}

The abundances for HD~219134, which is located 6.55 pc from the Sun, were measured by a total of 9 groups within the literature. The maximum [Fe/H] determination was by \citet{Mishenina15} with 0.84 dex, while Ramirez et al. (2007) measured a minimum of 0.04 dex and the median was 0.13 dex. Only three groups measured [Mg/H], where we use the median value of 0.32 dex and spread of 0.28 dex. The [Si/H] ratio spanned 0.04 dex per Valenti \& Fischer (2005) to 0.27 dex (Thevenin 1998, Thevenin \& Idiart 1999), with a median of 0.12 dex between the five contributing groups. The sodium, aluminum, and calcium abundance ratios were measured by three groups or less. The median for [Na/H] is 0.19 dex and [Al/H] is 0.23 dex, both with a spread of 0.13 dex. The [Ca/H] measurements had a smaller spread with 0.07 dex, while the median was 0.21 dex.

A total of 18 groups measured element abundances for 55 Cnc, which is 12.34 pc from the Sun. The [Fe/H] measurements ranged from 0.29 dex (Mishenina et al. 2004) to 0.67 dex (Takeda et al. 2007), with a median of 0.44 dex. Of the 18 groups, six of them measured [Mg/H], where the minimum determination was 0.34 dex (Delgado Mena et al. 2010), the maximum was 0.80 dex \citep{AllendePrieto}, and the median was 0.41 dex. The [Si/H] ratio was determined by 11 measurements and spanned 0.29 dex \citep{Gonzalez} to 0.57 dex (Allende Prieto et al. 2004), with a median of 0.40 dex. The sodium abundance ratio varied from 0.23 dex \citep{Gonzalez} to 0.63 dex per Shi et al. (2004) out of eight datasets, where the median was 0.49 dex. The [Al/H] determinations were only measured by five groups who ranged from 0.43 dex (Kang et al. 2011) to 0.55 dex (Luck \& Heiter 2005) and a median of 0.51 dex. Finally, [Ca/H] was measured by 7 groups, with a minimum of 0.09 dex (Zhao et al. 2002) to a maximum of 0.46 dex (Luck \& Heiter 2005); the median is 0.24 dex. 

HD~97658 is a star 21.11pc away from the Sun and was spectroscopically observed by six different groups. The variation in [Fe/H] goes from -0.05 (Petigura et al. 2011) to -0.28 dex (Mishenina et al. 2008) with a median value of -0.13 dex. The [Mg/H] ratio was only measured by two groups: Mishenina et al. (2008) and Mortier et al. (2013) who determined -0.18 dex and -0.27 dex, respectively. Three groups measured [Si/H] in HD~97658 and had a variation spanning -0.20 dex (Valenti \& Fischer 2005) to -0.28 dex (Mortier et al. 2013), with a median of -0.21 dex. Both Mortier et al. (2013) and Valenti \& Fischer (2005) determined [Na/H], citing -0.24 dex and -0.26 dex, respectively. Finally, only Mortier et al. (2013) measured [Al/H] = -0.23 dex and [Ca/H] = -0.25 dex.

For the host stars HD~219134, 55~Cnc, and HD~97658, we adopt three scenarios for their abundances: a case representing the median estimate from different studies (V0), an iron-rich (V1) and an iron-poor (V2) case. Whenever unspecified, the reference scenario V0 is used. For the three cases, we compile stellar abundance estimates in Table \ref{Vversion} based on Table \ref{stardata}. These stellar abundance estimates serve as proxies for the planet bulk abundance (see Table \ref{tabledata}, columns 3--7).  In section \ref{sensi} we examine how the spread of bulk abundance constraints influence our predictions on the planetary interiors of HD~219134b, 55~Cnc~e, and HD~97658b.

\setlength{\tabcolsep}{6pt}
\begin{table*}[ht]
\caption{Compiled stellar abundances representing median estimate from different studies (V0), an iron-rich (V1) and an iron-poor (V2) case. The unit is dex. \label{Vversion}}
\begin{center}
\begin{tabular}{l|ccc|ccc|ccc}
\hline\noalign{\smallskip}
\multirow{2}{*}{parameter}  &  \multicolumn{3}{c}{HD~219134 }&  \multicolumn{3}{c}{55 Cnc}&  \multicolumn{3}{c}{HD~97658} \\
& V0&V1&V2& V0&V1&V2& V0&V1&V2 \\
\hline\noalign{\smallskip}
$[\rm Fe/H]$ & 0.13 & 0.84&0.04&0.44 &0.67 & 0.29& -0.13&-0.05&-0.28\\
$[\rm Mg/H]$ & 0.32&0.09 &0.37&0.41 & 0.34&0.8 & -0.22&-0.27&-0.18\\
$[\rm Si/H]$& 0.12& 0.04&0.27& 0.40& 0.29& 0.57& -0.21&-0.28&-0.2\\
$[\rm Na/H]$ & 0.19& 0.17&0.32&0.49 & 0.23&0.63 &-0.25 &-0.24&-0.26\\
$[\rm Al/H]$& 0.23&0.16 &0.29&0.51 & 0.43& 0.55&-0.23 &-0.23&-0.23\\
$[\rm Ca/H]$& 0.21& 0.18&0.25&0.24 & 0.09& 0.46&-0.25 &-0.25&-0.25\\
\hline
\end{tabular}
\end{center}
\end{table*}

 \begin{table*}[ht] \setlength{\tabcolsep}{8pt}
 \caption{Data of selected exoplanets. V0 represents median abundance estimates, whereas V1 and V2 refer to iron-rich and iron-poor cases as listed in Table \ref{Vversion}; indices $\rm L$ and $\rm S$ of 55~Cnc~e refer to different radius estimates (see section \ref{prev}). 
 \label{tabledata}}
 \widesplit{%
\begin{tabular}{>{\bfseries}l|*{11}{l}}
\hline\noalign{\smallskip}
name 
\row{& $M$ [\ME] & $R$ [\RE]  & $\fesi$ & $\mgsi$ & Na$_2$O [wt\%] & Al$_2$O$_3$ [wt\%]   }{&  CaO [wt\%]  & $R_{\rm star}$ [\RS]  & $T_{\rm star}$ [T$_{\rm sun}$]  & semi-major axis [AU] }\\
\hline\noalign{\smallskip}
HD~219134 b (V0)
\row{& 4.36 $\pm$ 0.44 & 1.606 $\pm$ 0.086      &1.73$\pm$1.55  & 1.44$\pm$0.91 & 0.021  & 0.055  }{& 0.021 &    0.778 & 0.815& 0.038 }\\
HD~219134 b V1
\row{& 4.36 $\pm$ 0.44 & 1.606 $\pm$ 0.086      &10.68$\pm$1.55  & 1.02$\pm$0.91 & 0.01  & 0.023 }{& 0.01 &    0.778 & 0.815& 0.038 }\\
HD~219134 b V2
\row{& 4.36 $\pm$ 0.44 & 1.606 $\pm$ 0.086      &1.00$\pm$1.55  & 1.14$\pm$0.91 & 0.025  & 0.057  }{& 0.021 &    0.778 & 0.815& 0.038 }\\
Kepler 10-b
\row{& 3.7 $\pm$ 0.43 & 1.47 $\pm$ 0.02 & 2.18$\pm$0.41 & 1.14$\pm$0.28 & 0.02 & 0.05  }{& 0.02  &  1.065 & 0.978  & 0.017}\\
Kepler 93-b
\row{& 4.02 $\pm$ 0.68 & 1.478 $\pm$ 0.019        & 2.87$\pm$0.98 & 1.02$\pm$0.38 & 0.02 & 0.05  }{& 0.02 &   0.919&  0.983 & 0.054}\\
CoRoT 7-b
\row{& 4.386 $\pm$ 0.985 & 1.614 $\pm$ 0.102    &2.34$\pm$0.68  & 0.87$\pm$0.25  & 0.02 & 0.05  }{& 0.02  &  0.870 & 0.921  & 0.017}\\
HD~97658b (V0)
\row{& 7.866 $\pm$ 0.73 & 2.340 $\pm$ 0.165      &2.03$\pm$1.17  & 0.87$\pm$0.33 & 0.018  & 0.046  }{& 0.017 &    0.703 & 0.887& 0.08}\\
HD~97658b V1
\row{& 7.866 $\pm$ 0.73 & 2.340 $\pm$ 0.165      &2.87$\pm$1.17  & 0.93$\pm$0.33 & 0.018  & 0.045  }{& 0.017  &    0.703 & 0.887& 0.08}\\
HD~97658b V2
\row{& 7.866 $\pm$ 0.73 & 2.340 $\pm$ 0.165      &1.41$\pm$1.17  & 0.95$\pm$0.33 & 0.019  & 0.05  }{& 0.019&    0.703 & 0.887& 0.08}\\
55~Cnc~e$^{\rm L}$ (V0)
\row{& 8.37 $\pm$ 0.38 & 2.08 $\pm$ 0.16 &1.86$\pm$1.49 & 0.93$\pm$0.77 & 0.024  & 0.062  }{& 0.013  &  0.943 & 0.901  & 0.015}\\
55~Cnc~e$^{\rm S}$ (V0)
\row{& 8.37 $\pm$ 0.38 & 1.75 $\pm$ 0.13  &1.86$\pm$1.49  & 0.93$\pm$0.77 & 0.024  & 0.062  }{& 0.013  &  0.943 & 0.901  & 0.015}\\
55~Cnc~e$^{\rm S/L} $V1
\row{& 8.37 $\pm$ 0.38 & 1.75 $\pm$ 0.13 / 2.08 $\pm$ 0.16  &4.06$\pm$1.49  & 1.02$\pm$0.77 & 0.012  & 0.046  }{& 0.008 &  0.943 & 0.901  & 0.015}\\
55~Cnc~e$^{\rm S/L}$ V2
\row{& 8.37 $\pm$ 0.38 & 1.75 $\pm$ 0.13 / 2.08 $\pm$ 0.16   &0.89$\pm$1.49  & 1.54$\pm$0.77 & 0.024  & 0.048  }{& 0.016  &  0.943 & 0.901  & 0.015}
\end{tabular}%
}
\end{table*}

\subsection{Data}
\label{data}
The data \dat that we rely on to constrain planetary interiors (see section \ref{inversion}) are:
\begin{itemize}\itemsep0pt
\item planetary mass $M$,
\item planetary radius $R$,
\item bulk planetary ratio $\fesi$,
\item bulk planetary ratio $\mgsi$,
\item mantle composition of minor elements: CaO, Al$_2$O$_3$, Na$_2$O,
\item semi-major axis,
\item stellar radius $R_{\rm star}$,
\item stellar effective temperature $T_{\rm star}$.
\end{itemize}
Table \ref{tabledata} lists all data for the six selected exoplanets. Figure \ref{MR} shows their masses and radii plotted against M-R-curves of idealized compositions. 

The $\fesi$ mass ratio relates the mass of iron to silicon for the entire planet (core and mantle). Since all magnesium and silicate are in the mantle, $\mgsi$ equals $\mgsima$. As stated earlier, we are using two stellar abundance proxies, that is direct and non-direct proxy (section \ref{bulky}).
Following \citet{dorn}, we fix absolute abundances of minor refractories in the mantle (oxides of Ca, Na, and Al) to the stellar proxy. {  The system of Na$_2$O--CaO--FeO--MgO--Al$_2$O$_3$--SiO$_2$ is chosen, because it is able to explain 99\% of Earth's mantle.} Whenever stellar proxies of these minor compounds are not available, we assume chondritic abundances from \citet{Lodders09}.  Semi-major axis, stellar radius, and stellar effective temperature are also fixed parameters.

\subsection{Model parametrization}
\label{parametrization}

\begin{figure}[ht]
\centering
 \includegraphics[width = .33\textwidth]{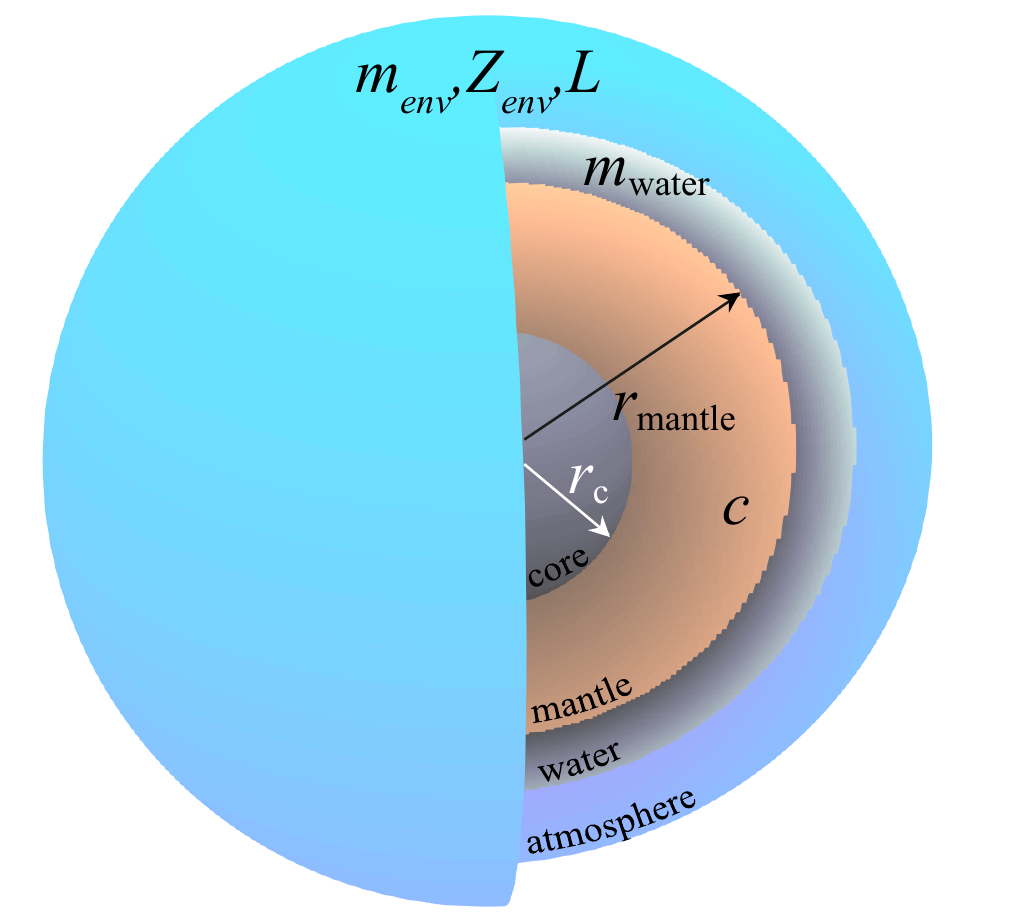}
 \caption{Illustration of model parametrization. Model parameters are core radius \rc, mantle composition $c$ comprising the oxides Na$_2$O--CaO--FeO--MgO--Al$_2$O$_3$--SiO$_2$, mantle radius \rsolid, mass of water \mice, mass of atmosphere \menv, atmosphere Luminosity $L$, and atmosphere metallicity \Zenv.  \label{Illustration}}
\end{figure}

Our exoplanet interior model consists of an iron core surrounded by a silicate mantle, a water layer and a gas layer as illustrated in Figure \ref{Illustration}. The key structural model parameters \m that we aim to constrain are:
\begin{itemize}\itemsep0pt
\item core radius \rc,
\item mantle $\fesima$,
\item mantle $\mgsima$,
\item mantle radius \rsolid,
\item mass of water \mice,
\item mass of atmosphere \menv,
 \item  atmosphere Luminosity $L$, 
\item  atmosphere metallicity \Zenv,
\end{itemize}
where \Zenv is defined as the atmosphere mass fraction of elements heavier than H and He (here C and O).

\subsection{Structure model}
\label{model}

Data \dat and model parameters  \m are linked by a physical model embodied by the forward operator $g(\cdot)$:

\begin{equation}
\dat = g(\m) \,.
\label{equ_forward}
\end{equation}
For a given model $\m$, the interior structure (density profile), total mass $M$, transit radius $R$, and $\fesi$ are computed for the purpose of comparing with observed data $\dat$.
The function $g({\bf m})$ is computed combining thermodynamic and Equation-of-State (EoS) modeling which is adopted from \citet{dornA}. In the following, we briefly summarize the detailed description given in \citet[][model I]{dornA}.

\paragraph{{\bf Iron core}}
For the core density profile, we use the EoS fit of iron in the hcp (hexagonal close-packed) structure provided by \citet{bouchet} on {\it ab initio} molecular dynamics simulations for pressures up to 1500 GPa. This EoS fit is optimal for the modeling of Earth-like exoplanets up to ten Earth masses.
The temperature gradient is assumed to be adiabatic.

\paragraph{{\bf Silicate mantle}}
We employ a self-consistent thermodynamic method within the Na$_2$O-CaO-FeO-MgO-Al$_2$O$_3$-SiO$_2$ model system. We assume thermodynamic equilibrium.  Equilibrium mineralogy and density are computed as a function of pressure, temperature, and bulk composition by Gibbs energy minimization \citep{connolly09}. For these calculations the pressure is obtained by integrating the load from the surface boundary, while temperature is obtained by integrating an Earth-like temperature gradient from the surface temperature of the mantle. This surface temperature equals the temperature at the bottom of the water layer.

\paragraph{{\bf Water layer}}
In order to compute the density profile of the water layer, we follow \citet{Vazan} using a quotidian equation of state (QEOS). This QEOS is in good agreement with widely used ANOES \citep{Thompson} and SESAME EoS \citep{Lyon}.
Above 44.3 GPa, we use the tabulated EoS from \citet{seager2007} that is derived from DFT simulations and predict a gradual transformation from ice VIII to X. We assume an adiabatic temperature profile.

\paragraph{{\bf Gas envelope and atmosphere model}}

For a given radius and mass of the solid interior, distance to the star $a$, stellar effective temperature $T_{\rm star}$, stellar radius $R_{\rm star}$, envelope luminosity $L$, gas metallicity \Zenv, and gas mass \menv, we solve the equations of hydrostatic equilibrium, mass conservation, and energy transport. As in \citet{VENTURINI2015}, we implement the CEA (Chemical Equilibrium with Applications) package \citep{CEA} for the EoS, which performs chemical equilibrium calculations for an arbitrary gaseous mixture,  including dissociation and ionisation, and assuming ideal gas behavior. We account for gas compositions of H, He, C and O which are key atmospheric elements.

An irradiated atmosphere is assumed at the top of the gaseous envelope, for which the analytic irradiation model of \citet{GUILLOT2010} is adopted. This irradiation model assumes a semi-gray, global temperature averaged profile. The temperature is defined by envelope luminosity $L$ and stellar irradiation. { \coco For the planets of interest, internal temperatures generally range between 1000 and 3000 K, intrinsic heat fluxes from the interior lie between 10 and 350 K.} Calibrated opacities for different equilibrium temperatures and solar metallicities from \citet{JIN2014} are used. When non-solar metallicities are considered (\Zenv $\ne$ 0.02), computed opacities are thus not self-consistent (see discussion in section \ref{Discussion}). 

The boundary between the irradiated atmosphere and the envelope is defined where the optical depth in visible wavelengths is much larger than 1 \citep{JIN2014}. Within the envelope, the usual Schwarzschild criterion is used to distinguish between convective and radiative layers. That is, if the adiabatic temperature gradient is larger than the radiative one, the layer is stable against convection.
In order to compute the planet radius of a model realization we evaluate where the chord optical depth becomes 2/3 \citep{GUILLOT2010}. This allows us to compare calculated planet radii with measured radii from primary transits.

\subsection{Prior information}
\label{prior}

Table \ref{tableprior} lists the ranges of prior information. Reasonable choices for the prior distributions are crucial such that we do not arbitrarily favor certain model realizations without justification. We have adopted the prior ranges from \citet{dornA}.

Prior bounds on $\fesima$ and $\mgsima$ are linked to the host star's photospheric abundances. Since all Si and Mg are assumed to be in the mantle, $\mgsistar$ defines the prior bounds on $\mgsima$, while $\mgsistar$ is Gaussian-distributed. The iron content, on the other hand, is distributed between core and mantle. Thus, the bulk constraint $\fesi$ defines only the upper bound of the prior on $\fesima$. There is also a numerical constraint that the absolute iron oxide abundance in the mantle cannot exceed 70~\%. For \menv and $L$, we assume a Jeffrey's prior, i.e., the logarithm of the parameter is uniformly distributed.

\begin{table}[ht]
\caption{Prior ranges of model parameters. \label{tableprior}}
\begin{center}
\begin{tabular}{ll}
\hline\noalign{\smallskip}
parameter  & prior range \\
\noalign{\smallskip}
\hline\noalign{\smallskip}
\rc         & 0 -- $R$ (uniform in $r_{\rm core}^3$) \\
$\fesima$           & 0 -- $\fesistar$(uniform)\\
$\mgsima$           & $\mgsistar$ (Gaussian)\\
\rsolid & 0 -- 1.1~$R$ (uniform) \\
\mice    & 0 -- 0.99~$M$ (uniform) \\
\menv  & $10^{-12}$ \ME -- 0.9~$M$ (Jeffrey's prior)\\
$L$                & $10^{18} - 10^{23}$~erg/s (Jeffrey's prior) \\
\Zenv             & 0 -- 1 (uniform in \Zenv$^{-1}$) \\
\hline
\end{tabular} 
\end{center}
\end{table}

%%%%%%%%%%%%%%%%%%%%%%%%%%%%%%%%%%%%%%%%%%%%%%%%%%%%%%%%%%%%%%%%%%%%%%%%%%%%%%
\section{Results}
\label{Results}

We apply our method to the six selected exoplanets with the data listed in Table \ref{tabledata} and first assume the direct abundance proxy ($\fesi = \fesistar$ and $\mgsi = \mgsistar$) and best guess stellar abundance estimates. We calculated the posterior distribution of model parameters and analyzed 2-dimensional (2-D) correlations between all 8 model parameters. 
In the following, we highlight a few (2-D) parameter correlations: \rsolid, \rc, and $\fesima$ (Figure \ref{4RF}), \rsolid and \mice (Figure \ref{1RW}); \rc and  \mice (Figure \ref{2WM}), as well as \Zenv and atmospheric radius fraction $r_{\rm env}/R$ (Figure \ref{3ZW}).

The correlation between mantle composition and core size (Figure \ref{4RF}) is introduced by the bulk abundance constraint $\fesi$. For planets for which the uncertainty on $\fesistar$ is lowest (below 34~\%), namely Kepler-10b, Kepler-93b, and CoRoT-7b, an increase in core-size leads to lower mantle $\fesima$ in order to fit bulk abundance constraints. This is illustrated by the horizontal shift in color (Figure \ref{4RF}b-d) from yellow to blue.  Whenever $\fesistar$ has a large uncertainty (e.g., about 90~\%, 80~\%, and 60~\% for HD~219134b, 55 Cnc, and HD~97658, respectively), the negative correlation between mantle size and mantle $\fesima$ dominates. In these cases the color shifts from yellow to blue vertically.  This is because the dominating datum is mass. More specifically, the core size is less well constrained, because for example a decrease in mantle size can be compensated by increasing $\fesima$ in order to fit mass without affecting \rc. For HD~97658b, actually both trends are visible, yellow roughly shifts along a diagonal.
%Note that the absolute amount of iron in the mantle depends as well on $\mgsima$.

 The correlation between \rsolid and \mice is generally strong as depicted in Figure \ref{1RW}. As expected from comparisons to idealized mass-radius-curves (see Figure \ref{MR}), it is possible to explain the data for HD~219134b, Kepler-10b, Kepler-93b, CoRoT-7b, and 55~Cnc~e$^{\rm S}$ with a rocky interior and no significant volatile layer. The full posterior range of the water mass fraction, however, reveals that significant volatile compounds cannot be excluded by the data. The possible mass fraction of water,\mice, spans from 0--35~\% (0--17~\% to explain the data within 1-$\sigma$ uncertainty) depending on the planet. Thus, their probability to be Earth-like is low, also since posterior distributions are generally large.  Only for Kepler-10b the range of possible water mass fractions is restricted to maximum few percents.
For the least dense exoplanets 55~Cnc~e$^{\rm L}$ and HD~97658b, \mice can be as large as 0.7 $M$ and 0.95 $M$, respectively (0.3 $M$ and 0.7 $M$, respectively, to fit the data within 1-$\sigma$ uncertainty, blue points in Figure \ref{1RW}).

Also, the (negative) correlation between mantle size \rsolid and \mice for all planets (Figure \ref{1RW}) is much stronger than between core size \rc and \mice (Figure \ref{2WM}). More specifically, this is because the parameters that most strongly affect planet mass are \rsolid  and \mice. An increase in \rsolid can be compensated by decreasing \mice, while still fitting planet mass. However, relative core size is mostly determined by $\fesi$.
%For example, increasing \mice requires decreasing \rsolid and leaves a possible range of mantle compositions and core sizes that are conditioned to $\fesi$. 

{ Figure \ref{3ZW} indicates that the correlation between gas metallicity and atmospheric radius fraction is generally weakly negative. 
The upper bound on \menv positively correlates with \Zenv (not shown)}. This is because the  same atmospheric thickness of a more metal-rich gas can only be obtained by increasing atmospheric mass. Among the six exoplanets, the upper bound on \menv differs depending on bulk density and data uncertainty. Thus the five potentially rocky exoplanets (Figure \ref{3ZW} a-e) can have an atmosphere of maximum mass of 10$^{-5}$-- 10$^{-7}$~\ME for low metallicities (\Zenv$~<~$0.2). For high metallicities (\Zenv$~>~$0.8), the maximum mass can be four magnitudes higher. In comparison the Earth's atmosphere has similar mass fraction of about $9\times10^{-7}$~\ME. For the volatile-rich exoplanets (Figure \ref{3ZW} e-f), {  the correlation between the atmospheric radius fraction $r_{\rm env}/R$ and \Zenv is strongest and the correlation between the upper bound on \menv and \Zenv seems to be negative}. Whether the planets could retain the constrained atmospheres with regard to possible photo-evaporative mass loss is not addressed in our model, but needs some post-analysis consideration (section \ref{evoloss}). 

Overall, correlations are generally weak since the problem addressed here is highly degenerate. In other words, an arbitrary change in 2 model parameters can be compensated by the remaining 6 model parameters in order to fit data. We find significant correlations between those model parameters that most strongly influence mass and/or radius and that countervail each other regarding this datum (e.g., \rsolid  and \mice both most strongly affect planet mass).

Finally, the predicted ranges for structural parameters strongly depend on the specific planet data and associated uncertainties. Even though we see a distinction between the potentially rocky (HD~219134b, Kepler-10b, Kepler-93b, CoRoT-7b, 55~Cnc~e$^{\rm S}$) and volatile-rich planet types (55~Cnc~e$^{\rm L}$ and HD~97658b), the distributions can vary significantly among planets of the same type. Among all studied planets, Kepler-10b seems to be best constrained and its interior is most similar to an Earth-like structure. It is our contention that a case-by-case analysis is essential for a thorough characterization of planet interiors. A comparison of previous results (section \ref{prev}) with our analysis shows very good agreement.

\subsection{Evaporative mass loss}
\label{evoloss}
Highly irradiated planets can lose large amounts of their atmospheres through hydrodynamic mass loss driven by the extreme ultraviolet and X-ray heating from their host stars. In order to investigate whether our data-constrained interior structures are stable in the sense that atmospheres are stable against photo-evaporation, we use simplistic analytical relationships  by \citet{lopez} and \citet{JIN2014}. A thorough analysis involving state-of-the-art mass loss models that couple hydrodynamic evaporation and thermal evolution models {\citep[e.g.][]{valencia10}} lies outside the scope of this study. Both simplistic scaling relations were fit against a large number of model runs and qualitatively describe the importance of mass loss but do not make detailed predictions. %It is important to note that these relations are valuable in understanding the significance of evaporative mass loss on our large populations of planet realizations. 
\citet{lopez} provide a fit for threshold fluxes at which significant atmospheric loss occurs as a function of core mass and
mass-loss efficiency for {Earth-like, rocky interiors and H/He atmospheres (of up to 60~\% in mass fraction). 
Similarly, \citet{JIN2014} provide a simplistic mass-radius-threshold below which planets have likely lost their entire H/He atmosphere. The application of both relationships to our posterior realizations suggest that all studied planets experience significant mass loss, except for HD~97658b (and mass-loss efficiencies below 0.05 regarding the \citet{lopez} relationship). These findings are also in agreement with previous studies, e.g., namely with the suggestions of \citet{valencia10} that CoRoT-7b would likely not possess a significant atmosphere as well as \citet{demory16} that 55~Cnc~e can be devoid of atmosphere.

 %Nonthermal mass-loss driven by stellar wind plasma flow is not discussed here since other aspects like intrinsic magnetic fields may protect a planetary atmosphere from nonthermal mass-loss.

In general, surface temperatures of the studied planets are high ($\sim$ 2000~K, depending on structure and planet case) which may imply molten rocks at the surface and outgassing from the planetary interior. Outgassing is a likely source for secondary atmospheres \citep[e.g.,][]{Schafer}. However, due to the high irradiation, significant mass loss may remove most of such atmospheres. Detailed models are needed to determine how much vapor could remain in equilibrium with possible rates of outgassing and evaporative mass loss.

\begin{figure*}[]
\centering
 \includegraphics[width = .9\textwidth, trim = 0cm 0cm 0cm 0cm, clip]{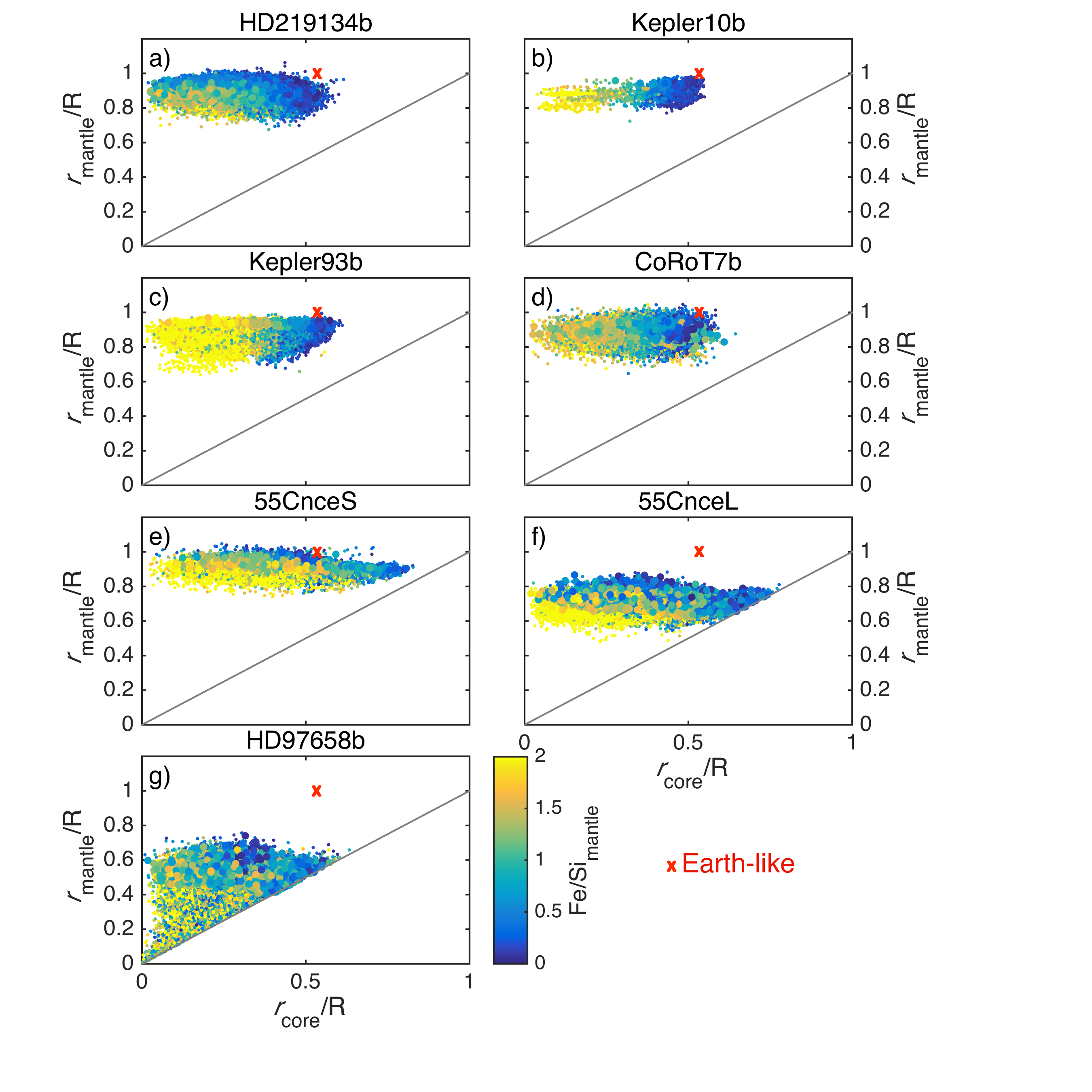}
 \caption{Sampled two-dimensional (2-D) marginal posterior pdfs for the six selected exoplanets (a-f) showing the correlation between core and mantle sizes, \rc, and \rsolid and mantle $\fesima$. Bigger dots explain the data within 1-$\sigma$ uncertainty. The straight line describes the lower limit of the mantle size where \rsolid = \rc. Earth-like parameters are depicted by the red cross.  \label{4RF}}

\end{figure*}

\begin{figure*}[]
\centering
 \includegraphics[width = .9\textwidth, trim = 0cm 0cm 0cm 0cm, clip]{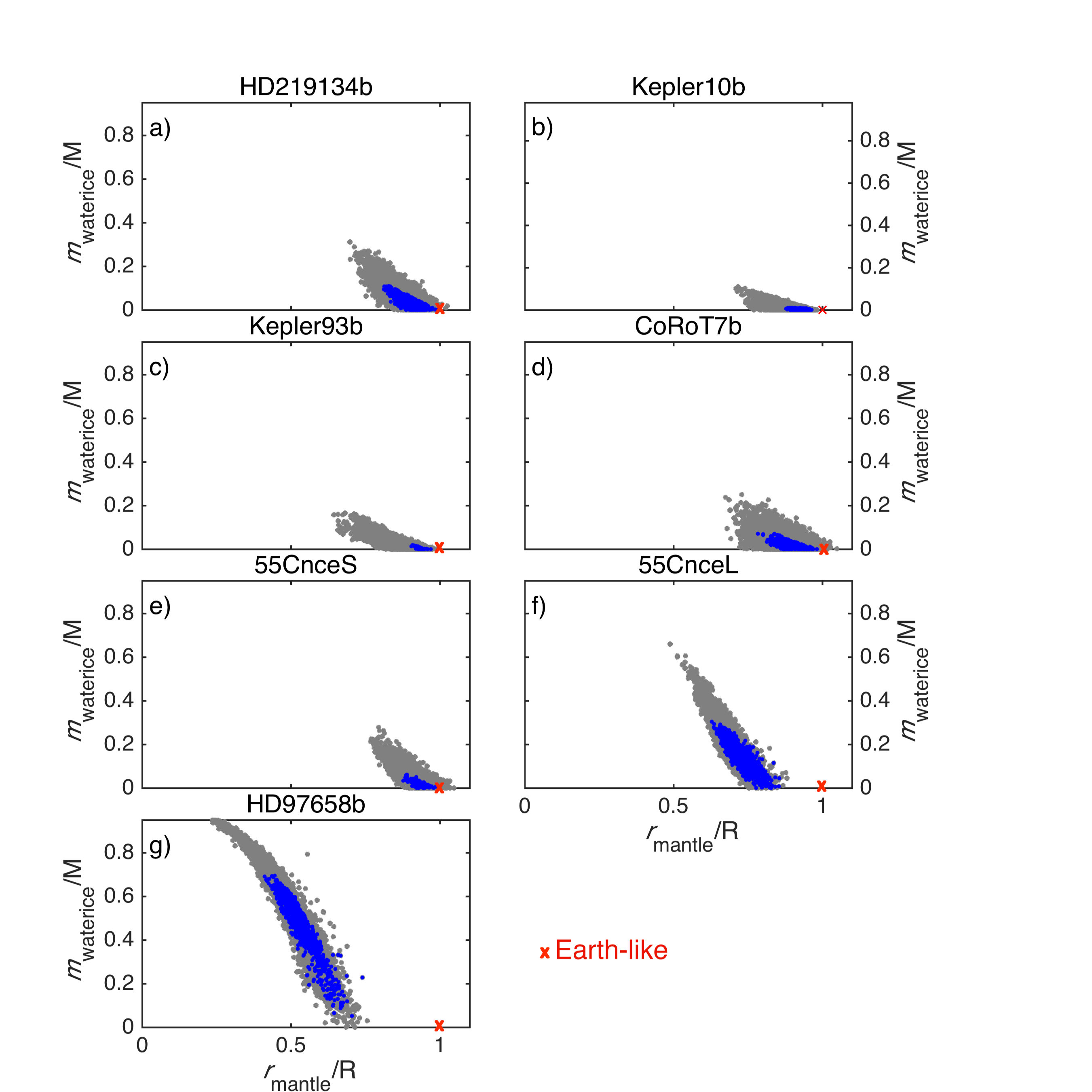}
 \caption{Sampled 2-D marginal posterior pdfs for the six selected exoplanets (a-f) showing the correlation between mantle size \rsolid and mass of water \mice. Blue points explain the data within 1-$\sigma$ uncertainty. Earth-like parameters are depicted by the red cross.\label{1RW}}
 
\end{figure*}

\begin{figure*}[]
\centering
 \includegraphics[width = .9\textwidth, trim = 0cm 0cm 0cm 0cm, clip]{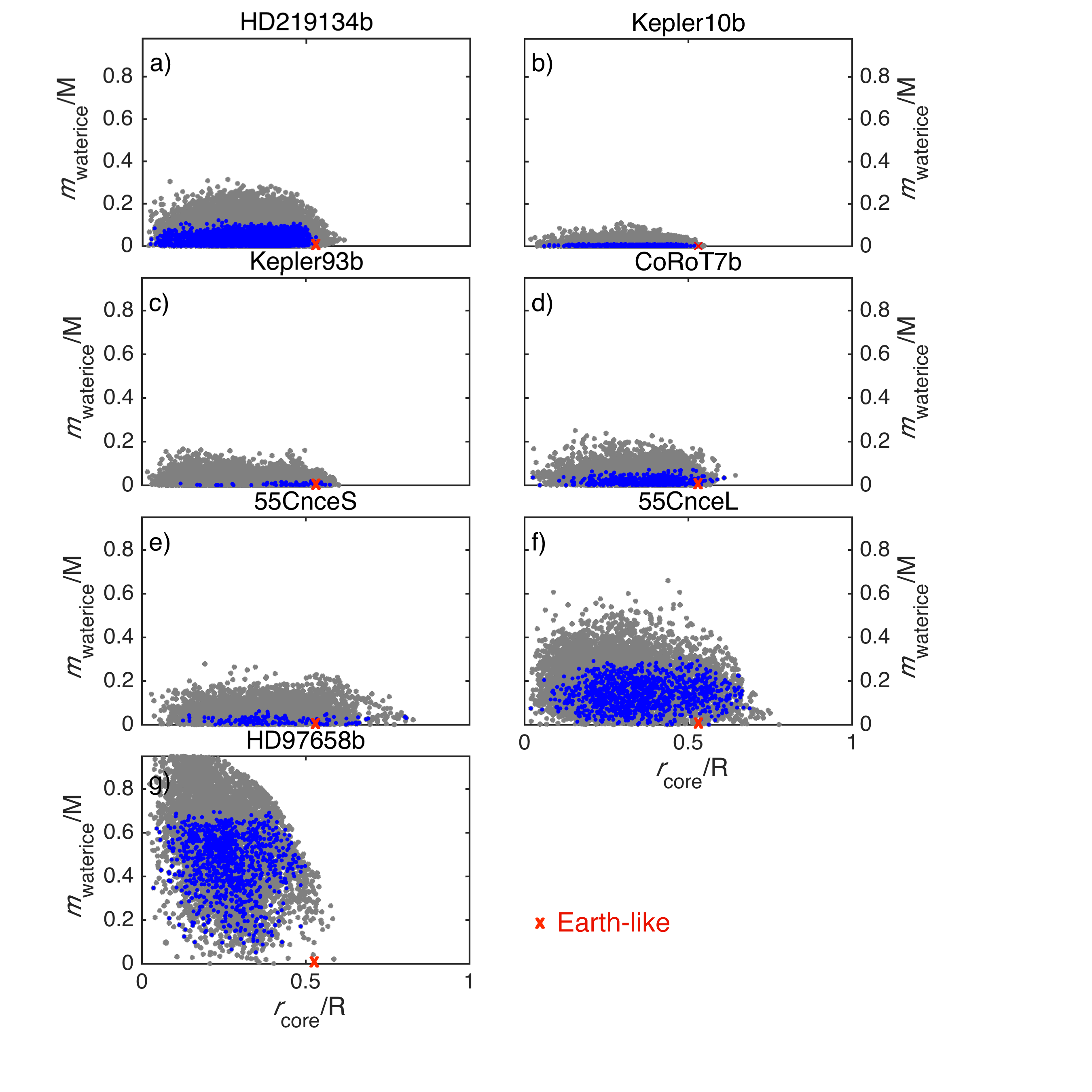}
 \caption{Sampled 2-D marginal posterior pdfs for the six selected exoplanets (a-f) showing the correlation between core size \rc and mass of water \mice. Blue points explain the data within 1-$\sigma$ uncertainty. Earth-like parameters are depicted by the red cross. \label{2WM}}

\end{figure*}

\begin{figure*}[]
\centering
 \includegraphics[width = .9\textwidth, trim = 0cm 0cm 0cm 0cm, clip]{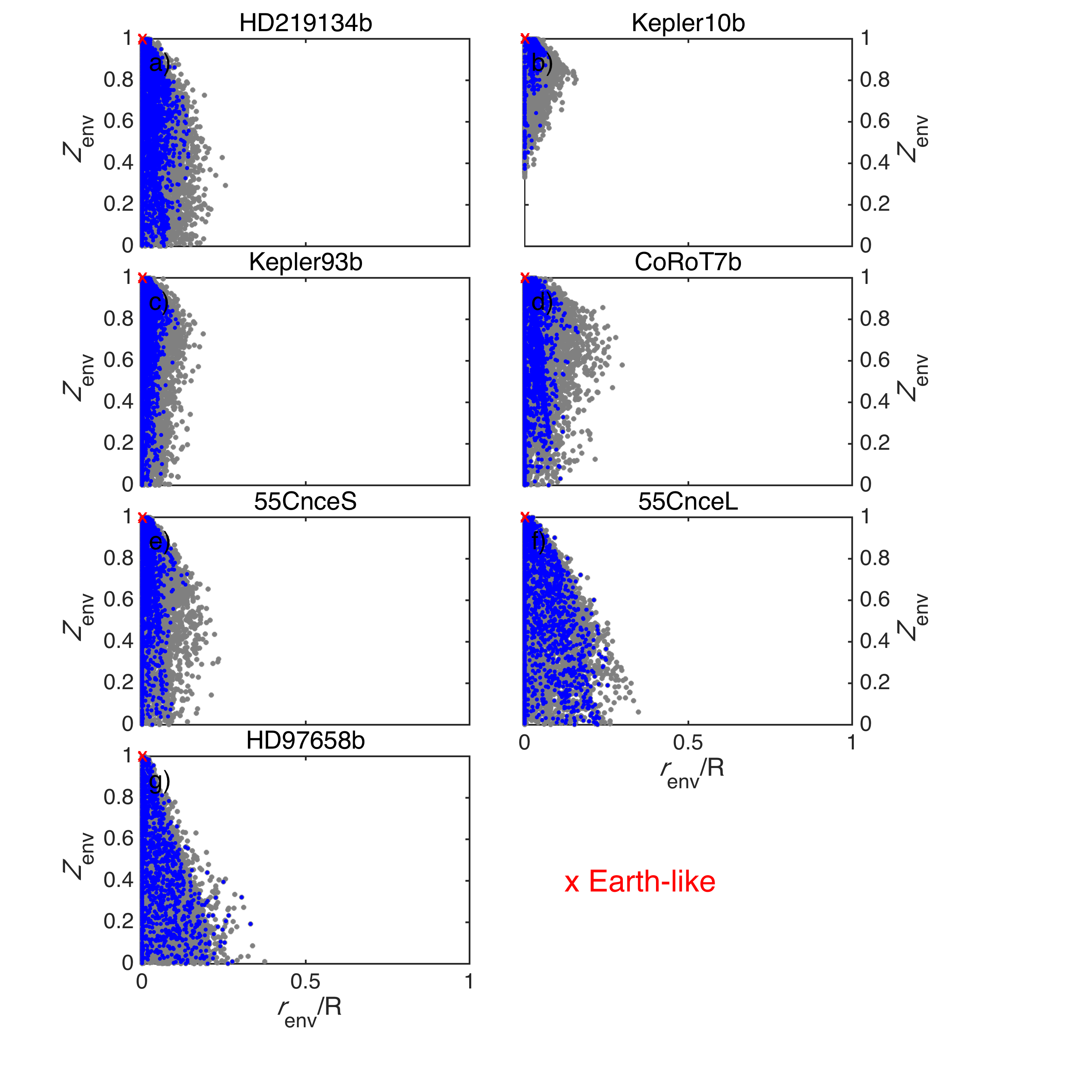}
 \caption{Sampled 2-D marginal posterior pdfs for the six selected exoplanets (a-f) showing the correlation between gas metallicity \Zenv and atmospheric radius fraction ${\it r}_{env}/R$. Blue points explain the data within 1-$\sigma$ uncertainty. Earth-like parameters are depicted by the red cross. \label{3ZW}}

\end{figure*}

%_______________________________________________________
\subsection{Sensitivity of structural parameters to bulk abundance constraints}
\label{sensi}
Here, we examine the sensitivity of interior predictions on the spread of planetary bulk abundance constraints. We do this by testing how interior predictions for HD~219134b, 55~Cnc~e, and HD~97658b vary in light of different bulk abundance estimates taken from: (I) best guess stellar abundance estimates, (II) iron-rich (V1), and (III) iron-poor (V2) abundance estimates (Table \ref{tabledata}). Results are shown in Figure \ref{Sensitivity}.  
The blue curves for the best guess abundance estimate refer to the same posterior pdfs previously shown in Figures \ref{4RF}--\ref{3ZW}.

The spread in interior parameters is largest for HD~219134b and smallest for HD~97658b. This is because the difference in bulk abundance constraints between V1 and V2 is a factor of 10 for HD~219134b, whereas for 55~Cnc~e and HD~97658b the difference is only a factor of 4.5 and two, respectively.
Except for $\fesima$ and \rc, differences between the {cumulative distribution functions (cdfs)} are generally minor. Compared to V2, the 50th-percentile in \rc and $\fesima$ increases by up to 20~\% and 200~\% for V1, respectively, for \rsolid and \mice these variations are on the order of a few percents. Differences in \menv are minor.

The differences in core size and mantle composition introduced by the use of different stellar abundance estimates are even more pronounced  when reduced uncertainties for mass and radius are available. This will be addressed in future parameter studies.

\begin{figure*}[]
\centering
  \includegraphics[width = 1.\textwidth]{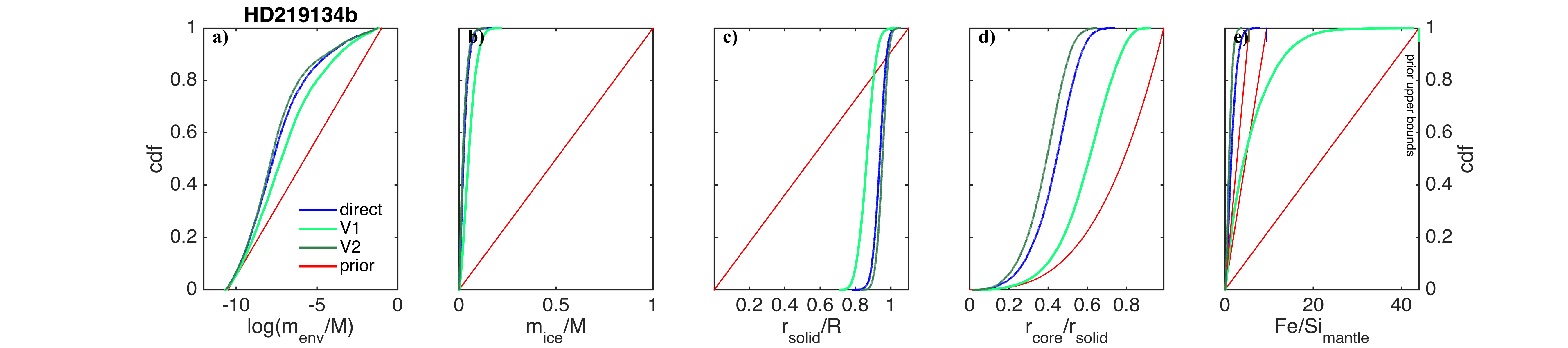}\\
 \includegraphics[width = 1.\textwidth]{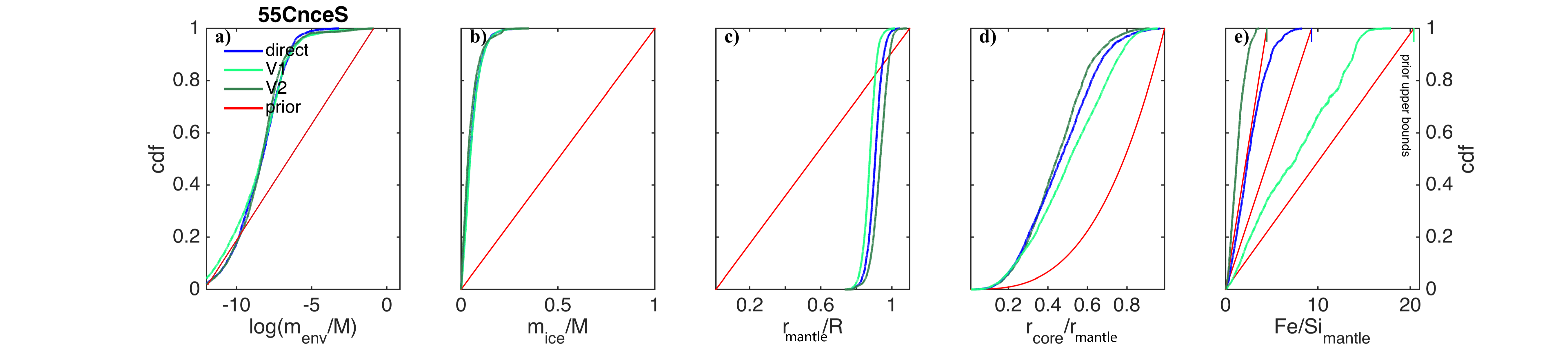}\\
  \includegraphics[width = 1.\textwidth]{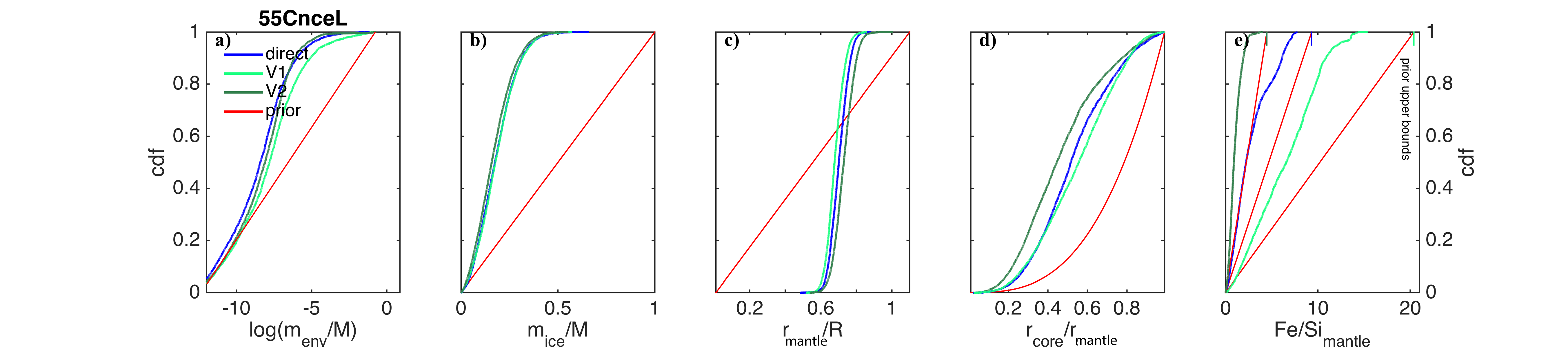}\\
   \includegraphics[width = 1.\textwidth]{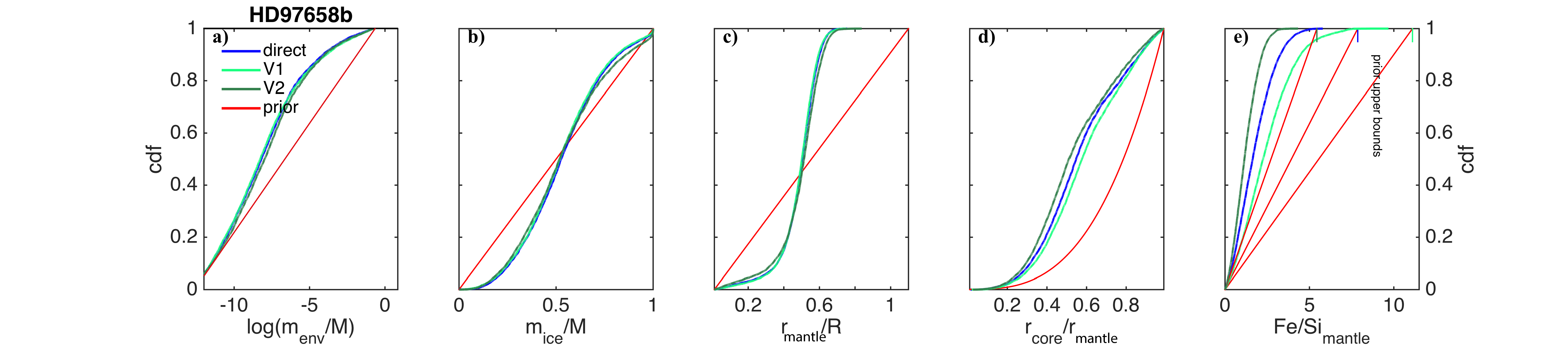}\\
 \caption{{Application of V0, V1, and V2 stellar abundance estimates:  cdfs} of sampled 1D marginal posterior (green to blue) and prior (red) for HD~219134b, 55~Cnc~e$^{\rm S}$, 55~Cnc~e$^{\rm L}$, and HD~97658b: (a)  \mice, (c) \rsolid, (d) \rc, and (e) $\fesima$. Posterior cdfs refer to the different host star abundance estimates (V1- iron rich, V2 - iron poor) listed in Table \ref{tabledata}. The dashed lines are the 95~\% lower and upper confidence bounds, which almost completely overlap the solid lines. Depending on the stellar abundance proxy estimates, the upper prior bound in (e) differs, which is indicated by the vertical colored lines corresponding to the respective proxy. \label{Sensitivity}}

\end{figure*}

%__________________weird correlation________________________
%_______________________________________________________

\subsection{Direct versus non-direct stellar abundance proxy}
% 5 planets with 1:weird correlation
\label{nondirect}
Here, we investigate the influence of different stellar abundance proxies by comparing retrieved interior structures obtained with the direct ($\fesi~=~\fesistar$ and $\mgsi~=~\mgsistar$) and non-direct abundance proxy ($\fesi~=~10^{-5}~\fesistar$ and $\mgsi =~0.5~\mgsistar$) (see section \ref{bulky}) on best guess stellar abundance estimates. The non-direct abundance proxy implies an extremely low bulk iron  (nearly iron-free) content of the planets.  With the constraint of a nearly iron-free planet, mass and radius of Kepler-93b and Kepler-10b can not be fit within 1-$\sigma$ uncertainty. All other studied planets can be explained within 1-$\sigma$ uncertainty, because data uncertainty is simply large enough or the bulk density (e.g., 55~Cnc~e$^{\rm L}$ and HD~97658b) can be explained without a large iron core.

Figure \ref{cdf2} illustrates the differences in parameter estimates for the direct (blue curves) and non-direct (green curves) abundance proxy. Again, the blue curve refer to the same posterior pdfs previously shown in Figures \ref{4RF}--\ref{3ZW}, and Figure \ref{Sensitivity} (blue curve).

The obtained structural parameters differ significantly between the two tested abundance proxies. As expected, using the non-direct abundance proxy compared to the direct abundance proxy leads to extremely low iron content in the mantle and tiny iron cores. In order to still fit mass, the size of rocky interior is larger.
The 50th-percentiles of the parameter cdfs in Figure \ref{cdf2} change as follows for the extreme sub-stellar bulk abundance constraints: 
\begin{itemize}\itemsep0pt
\item \rc decreases by 78\% - 83\%,
\item \rsolid increases by 2\% - 13\%,
\item $\fesima$ decreases by 100\%,
\item \mice decreases by 9\% - 50\%,
\end{itemize} 
Since structural variability in core and mantle is strongly restricted by the extremely low bulk abundance constraint, core and mantle parameters are generally better constrained (i.e., cdf curves are steeper).

With our methodology and the given data, such extreme sub-stellar abundances cannot be excluded for potentially rocky exoplanets, although extremely low iron content is disfavored for Kepler-93b and Kepler-10b. It is important to note that such low iron contents are very unlikely for volatile-rich planets, because accreted volatiles originate from primordial-disk regions beyond the ice-line at which also Mg-Fe-Si-bearing species condensed and likely reached the stellar ratio \citep{thiabaud}.

\begin{figure*}[]
\centering
{\includegraphics[width = .8\textwidth,height=0.135\textheight]{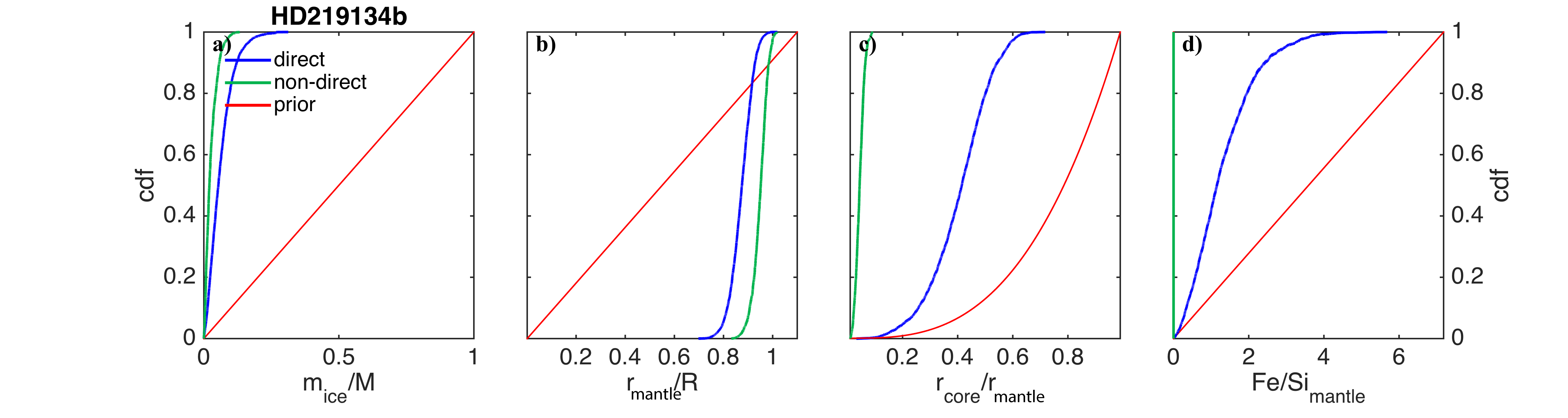}}\\
{\includegraphics[width = .8\textwidth,height=0.135\textheight]{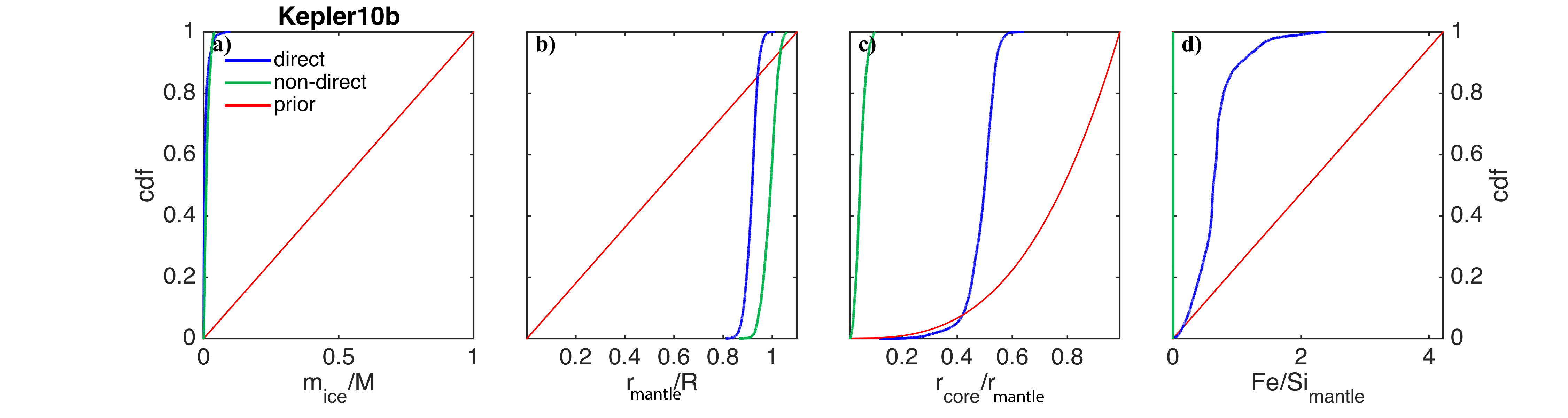}}\
{ \includegraphics[width = .8\textwidth,height=0.135\textheight]{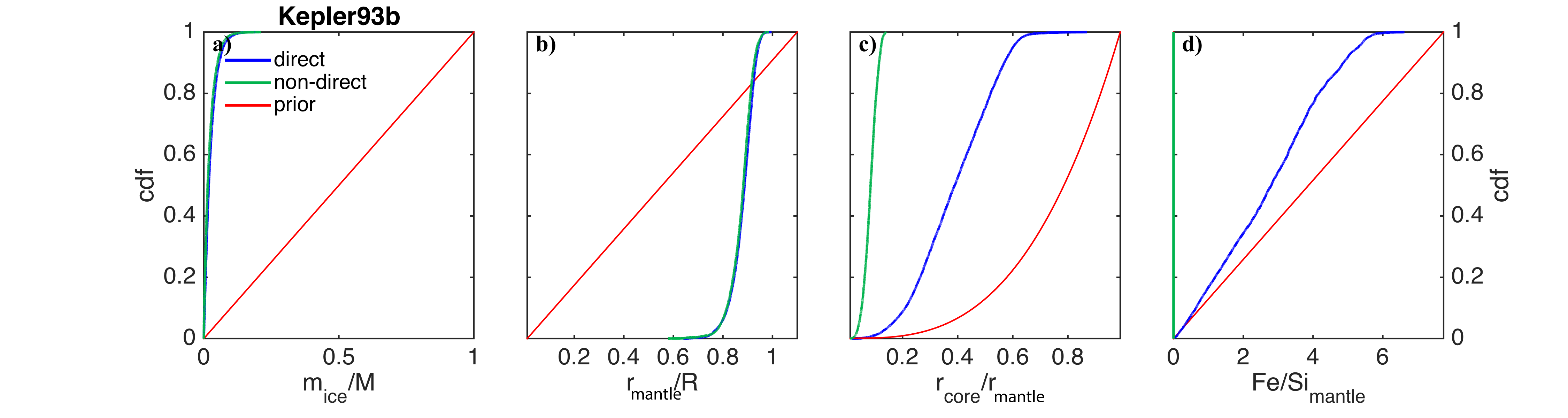}}\\
{  \includegraphics[width=.8\textwidth,height=0.135\textheight]{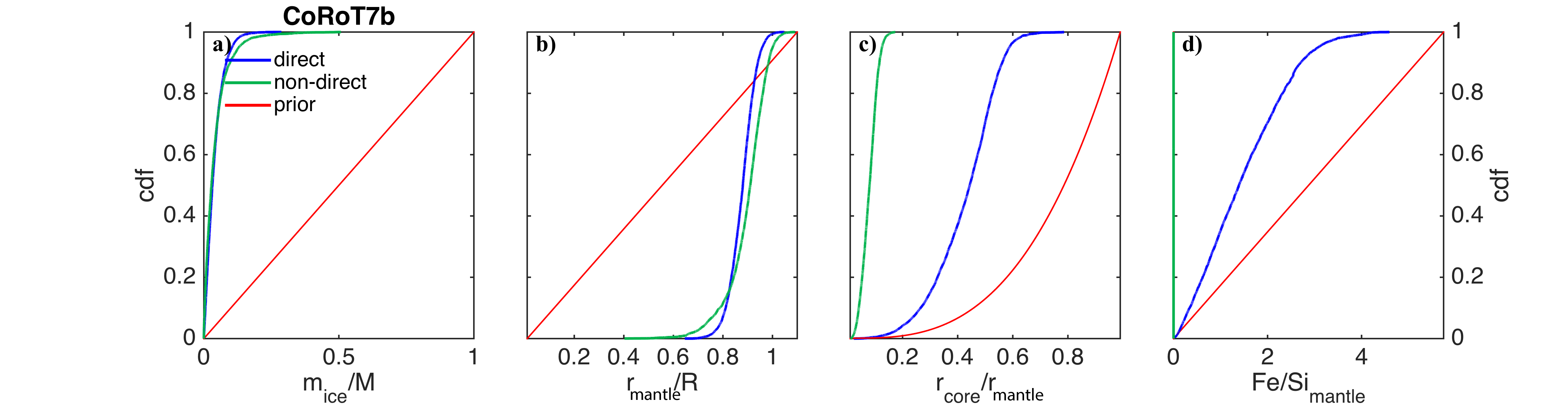}}\\
{ \includegraphics[width=.8\textwidth,height=0.135\textheight]{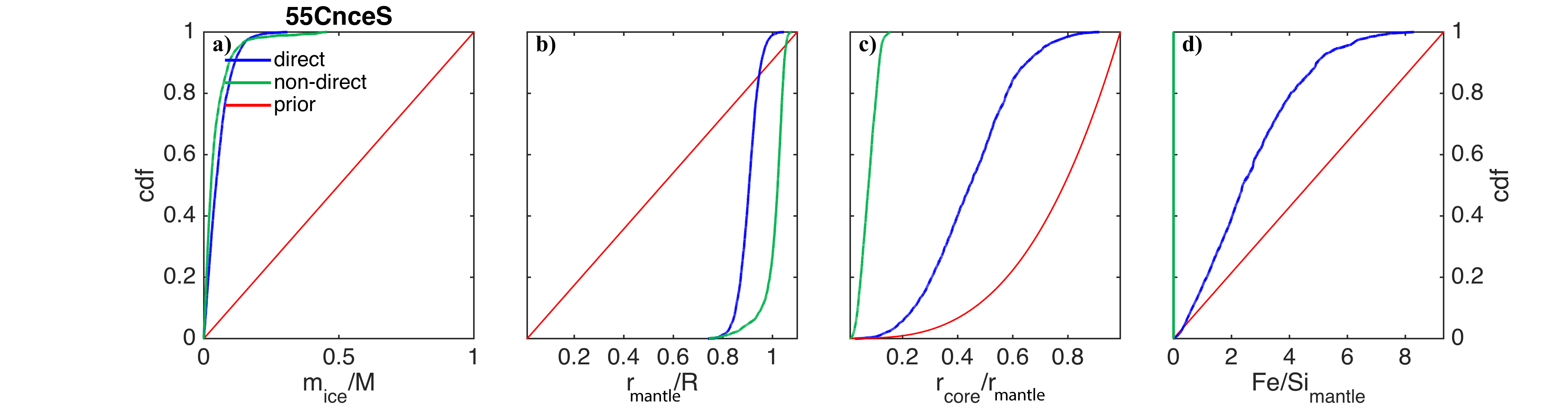}}\\
{ \includegraphics[width =.8\textwidth,height=0.125\textheight,height=0.135\textheight]{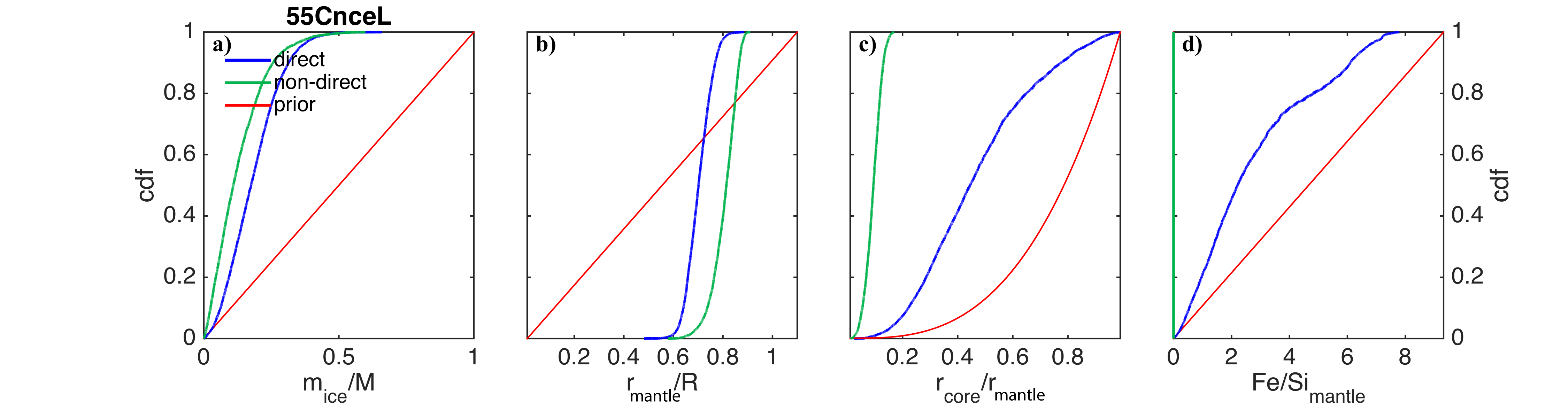}}\\
{   \includegraphics[width=.8\textwidth,height=0.135\textheight]{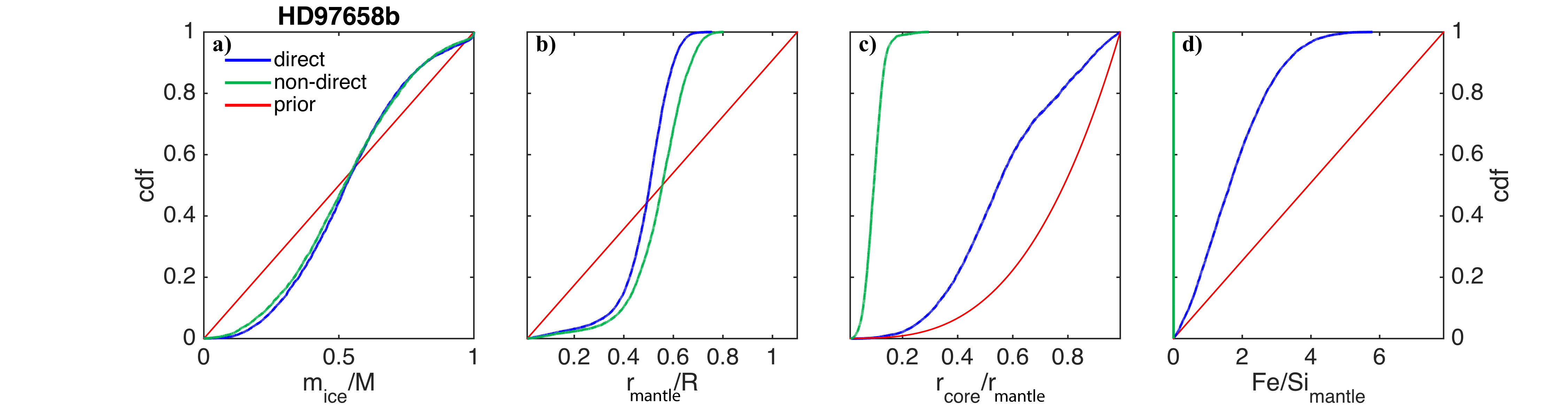}}\\
\caption{{Application of direct and non-direct abundance proxies: cdfs} of 1D marginal posterior pdfs and prior pdfs for HD~219134b, Kepler-10b, Kepler-93b, CoRoT-7b, 55~Cnc~e$^{\rm S}$, 55~Cnc~e$^{\rm L}$, and HD~97658b: (a) \menv, (b) \Zenv, (c) \mice, (d) \rsolid, (e) \rc. Prior (red), posterior pdf in blue and green for direct and non-direct stellar-planetary abundance correlation, respectively. See discussion in section \ref{nondirect}. \label{cdf2} }
\end{figure*}

%\subsubsection{Bayesian model comparison}
In order to quantify how much the direct abundance proxy is preferred over the non-direct abundance proxy, we use the posterior odds, $K$, that is a multiplication of Bayes factor and prior odds {\coco \citep{berger}}:

\begin{equation}
K = \frac{{\rm p}(\dat|H_{\rm d}) {\rm p}(H_{\rm d})}{{\rm p}(\dat|H_{\rm n}) {\rm p}(H_{\rm n})} \,\, .
\end{equation}
Here, $H_{\rm d}$ and $H_{\rm n}$ denote the hypothesis of validity of the direct and non-direct abundance proxy, respectively, and the probability

\begin{equation}
{\rm p}(\dat|H_{\rm d}) = \int{{\rm p}(\m|H_{\rm d}){\rm p}(\dat|\m,H_{\rm d}) d\m}\,,
\end{equation}
represents how well some data are explained under the assumption of a hypothesis $H_{\rm d}$.
Instead of the Bayes factor integral we use the maximum likelihood estimate. We obtain K-values around $10^1$, except for Kepler-93b and Kepler-10b with K-values of $10^{2}$. This means that only for Kepler-93b and Kepler-10b, there is a strong evidence that the direct abundance proxy better explains the planetary mass and radius. 

Generally, it will be very difficult to study the validity of a direct correlation of relative bulk abundances between star and hosted exoplanet given usual data uncertainties. Further studies on solar system objects are needed to improve our understanding of such abundance correlations and their dependence on planet formation history.

%%%%%%%%%%%%%%%%%%%%%%%%%%%%%%%%%%%%%%%%%%%%%%%%%%%%%%%%%%%%%%%%%%%%%%%%%%%%%%
\section{Discussion of the assumptions}
\label{Discussion}

The success of an inference method is subject to the limitations and assumptions of the forward function $g(\cdot)$. Firstly, we consider pure iron cores, which may lead to a systematic overestimation of core density and thus an underestimation of core size, if volatile compounds in the core are significant. The trade-off between predicted and independently inferred core sizes can be on the order of few to tens of percents, as shown for the terrestrial solar system planets in \citet{dorn}.

Secondly, our approach assumes sub-solidus conditions for core and mantle as well as a perfectly known EoS parameters for all considered compositions. For the six exoplanets considered here, their pressures and temperatures exceed the ranges for which these parameters can be measured in the laboratory. Clearly, extrapolations introduce uncertainty.

Thirdly, we have assumed pure water composition of the ice layer, but other compounds such as CO, CO$_2$, CH$_4$, NH$_3$, etc. are also possible since the additional elements are relatively abundant. We have used water as a proxy for these volatiles in general because 
(1) oxygen is more abundant than carbon and nitrogen in the universe, and (2) water condenses at higher temperatures than ammonia and methane. 
%Furthermore, density contrasts between these volatiles are much smaller compared to mantle silicates or gases. Therefore using pure water as a proxy for volatile compounds only artificially reduced degeneracy to a reasonable amount. %Also, note that water layers are in super-critical form, since the planets receive much of their heating from their central star. Thus the water would be in super-critical vapour form for most scenarios. For example, laboratory observations lead to the distinction of three fluid phases: (i) a molecular regime at low pressure, (ii) a nonmetallic ionic regime at intermediate pressure and temperature, and (iii) a metallic regime at high pressure and temperature \citep{Cavazzoni, French}. Here, we use a QEOS that treats water as a mixture of atoms \citep{Vazan}.

Fourthly, our atmospheric model uses prescribed opacities of \citet{JIN2014}, which is suitable for solar abundances. In this aspect, the prescription is therefore not self-consistent with the non-solar metallicity cases we present in this work. We have checked that using different infrared and/or visible opacities can introduce errors in radius of $\sim$5~\%. Models that compute line-by-line opacities with their corresponding atmospheric abundances should be performed in the future to compute planetary radii in a self-consistent way.
Another aspect is our assumption of ideal gas behavior. We have checked that for atmospheric mass fractions (\menv/$M$) of up to 1~\%, the difference in the radius between using the \citet{saumon} EOS and ideal gas (for H-He atmospheres) can reach 10~\%. For the vast majority of our results, atmospheric mass fractions are smaller than 0.1~\%, and thus the ideal gas assumption does not introduce a remarkable flaw. 
 
Since our interior models are static, evolutionary effects on the interior structure are not accounted for in the forward model. As demonstrated in section \ref{evoloss}, constraints on evolutionary processes such as thermal mass loss of gaseous layers can be roughly approximated by post-analysis considerations. We have used very approximate analytic scaling laws from \citet{lopez} and \citet{JIN2014} to rule out gas-rich planet realizations.

%The main point of our study is that the inferred interior realizations span the complete range of possible solutions. In addition, our sensitivity study on varying abundance constraints by a factor of 2-4 shows that a changes in structural parameters are generally negligible, except for \rc and $\fesima$; minor changes are also seen for \rsolid and \mice. Therefore, the common assumption of solar-like refractory abundances may introduce biases.

%%%%%%%%%%%%%%%%%%%%%%%%%%%%%%%%%%%%%%%%%%%%%%%%%%%%%%%%%%%%%%%%%%%%%%%%%%%%%%
\section{Summary}
\label{Conclusions}

We have presented the application of a generalized Bayesian inference method from \citet{dornA} in order to constrain the general interior structure of six selected super Earths: HD~219134b, Kepler-10b, Kepler-93b, CoRoT-7b, 55~Cnc~e, and HD~97658b. Their available data are mass, radius, and bulk abundance constraints on relative refractory elements (Fe, Mg, Si) measured in the photosphere of their host stars. We assume a general structure of core, mantle, ice and gas for which we constrain core size, mantle thickness and composition, mass of water, and key characteristics of the atmosphere (mass, intrinsic luminosity, composition).  We have investigated the sensitivity of the constrained interior parameters on the spread of bulk abundance estimates as determined by research groups with different techniques \citep{Hinkel14}. We have also determined how predictions of interior parameters are affected by a very non-direct correlation of relative abundances between star and planet, compared to a direct correlation.  Our findings are summarized as follows:

\begin{itemize}\itemsep2pt
\item The available data of the six studied exoplanets enables us to constrain interior structure and to determine parameter correlations. Depending on data and data uncertainty, the ability of constraining interior structure differs. Kepler-10b appears to be best constrained, since data uncertainties are relatively small.

\item In general, correlations between interior parameters are weak. We find strongest 2-D correlations between the size of the rocky interior \rsolid and water mass fraction \mice. 

\item We determine an upper range of possible atmospheric masses that positively correlate with gas metallicity. Whether the planets can retain the constrained atmospheres with regard to the high irradiation of their host stars is evaluated with simplistic analytical relationships that determine the importance of photo-evaporative mass loss.

\item The data of most of the studied exoplanets can not exclude massive amounts of water. Even for the potentially rocky exoplanets, the possible  range of water mass fraction can reach up to 35~\%, depending on the planet. Only for Kepler-10b, a massive water mass fraction can be excluded (\mice/$M~<~1$~\%).

\item Discrepancies in the stellar abundance estimates that are used as proxies for planetary bulk abundance mainly affect the prediction of mantle composition and core size. Effects on the other parameters of atmosphere, water layer, and size of rocky interior are small.

\item Extremely sub-stellar abundance constraints are less in agreement with mass and radius data for the potentially rocky exoplanets, but cannot be excluded without additional considerations. For instance, potentially volatile-rich exoplanets likely accreted material that originated from regions beyond the ice line, where Fe-Mg-Si bearing species reached stellar ratios. 
\end{itemize}

Available mass and radius data of exoplanets cannot be used to verify whether a correlation between stellar and planetary relative abundances of Fe, Mg, and Si differs from a direct 1:1 correlation. This is because degeneracy in interior structure of exoplanets is too large. More detailed studies of terrestrial solar system planets, e.g. Mars and Venus, would greatly help to improve our understanding of the discussed abundance constraints.

In order to rigorously characterize interior structure for an exoplanet, it is crucial to perform a case-by-case inference method, since constrained interior distributions highly depend on data and data uncertainties. The range of possible exoplanet interiors is large, since the problem is highly degenerate. Commonly used mass-radius-curves of idealized compositions for comparison against observed masses and radii can only serve as a rough planet characterization. {Bayesian inference methods offer a powerful way to determine exoplanet interiors sensibly.}

%\end{linenumbers}
%%%%%%%%%%%%%%%%%%%%%%%%%%%%%%%%%%%%%%%%%%%%%%%%%%%%%%%%%%%%%%%%%%%%%%%%%%%%%%
\begin{acknowledgements}

This work was supported in part by the Swiss National Foundation under grant 15-144. It was in part carried out within the frame of the National Centre for Competence in Research PlanetS. We would like to thank James Connolly for informed discussions, Amaury Thiabaud for making available the data published in \citet{thiabaud}. N.R.H would like to thank CHW3. The results reported herein benefited from collaborations and/or information exchange within NASA's Nexus for Exoplanet System Science (NExSS) research coordination network sponsored by NASA's Science Mission Directorate.

\end{acknowledgements}

\begin{appendix}  
\section{Element abundance notations}
\label{app}
Typically in astronomy, the total number of absorbing atoms
is scaled with respect to hydrogen (or 10$^{12}$ atoms), which is the
most common element in the universe. Geologists often use 10$^6$ silicon atoms as a scalar.
In addition, the abundance ratio within the star is referenced to
the same element abundance ratio within the Sun, the most well measured
star, such that the star can be defined to be rich, poor,
or solar-like with respect to that element. The unit of dex is described
in mathematical terms as:
\begin{equation}
[X/H]=\log(N_X/N_H)_{star} - \log(N_X/N_H)_{Sun}\,,
\end{equation}
where $N_X$ and $N_H$ are the number of element X and hydrogen
atoms per unit volume, respectively. When an abundance ratio
is given in square brackets, it denotes that it is in reference to
the Sun. The majority of our discussion is with respect to abundance
mass ratios without reference to the Sun, X/H (see Table
3, columns 3 and 4).
\end{appendix}

%%%%%%%%%%%%%%%%%%%%%%%%%%%%%%%%%%%%%%%%%%%%%%%%%%%%%%%%%%%%%%%%%%%%%%%%%%%%%%
%%%%%%%%%%%%%%%%%%%%%%%%%%%%%%%%%%%%%%%%%%%%%%%%%%%%%%%%%%%%%%%%%%%%%%%%%%%%%%
%%% REFERENCES %%%

\label{lastpage}

\end{document}